\documentclass[twocolumn]{aastex631}

\usepackage{amsmath}
\usepackage{amssymb}
\usepackage{graphicx}
\usepackage{gensymb}
\usepackage{booktabs}
\usepackage{diagbox}

\begin{document}
\title[LATRt Program]{Simons Observatory: Pre-deployment Performance of a Large Aperture Telescope Optics Tube in the 90 and 150~GHz Spectral Bands}
\correspondingauthor{Carlos E. Sierra}

\author[0000-0002-9246-5571]{Carlos E. Sierra}
\email{csierra@uchicago.edu}
\affiliation{Department of Physics, University of Chicago, Chicago, IL 60637, USA}
\affiliation{Kavli Institute for Cosmological Physics, University of Chicago, Chicago, IL 60637, USA}

\author[0000-0003-1248-9563]{Kathleen Harrington}
\affiliation{High Energy Physics Division, Argonne National Laboratory, Lemont, IL 60439, USA}
\affiliation{Department of Astronomy and Astrophysics, University of Chicago, Chicago, IL, 60637, USA}

\author[0000-0002-6971-8809]{Shreya Sutariya}
\affiliation{Department of Physics, University of Chicago, Chicago, IL 60637, USA}

\author[0000-0003-1942-1334]{Thomas Alford}
\affiliation{Department of Physics, University of Chicago, Chicago, IL 60637, USA}

\author[0000-0001-5374-1767]{Anna M. Kofman}
\affiliation{Department of Physics and Astronomy, University of Pennsylvania, Philadelphia, PA, 19104, USA}

\author{Grace E. Chesmore}
\affiliation{Department of Physics, University of Chicago, Chicago, IL 60637, USA}

\author[0000-0002-6338-0069]{Jason E. Austermann}
\affiliation{Quantum Sensors Division, NIST, Boulder, CO 80305, USA}

\author[0000-0002-7888-6222]{Andrew Bazarko}
\affiliation{Department of Physics, Princeton University, Princeton, NJ, 08544, USA}

\author[0000-0003-1263-6738]{James A. Beall}
\affiliation{Quantum Sensors Division, NIST, Boulder, CO 80305, USA}

\author[0000-0002-2971-1776]{Tanay Bhandarkar}
\affiliation{Department of Physics and Astronomy, University of Pennsylvania, Philadelphia, PA, 19104, USA}

\author[0000-0002-3169-9761]{Mark J. Devlin}
\affiliation{Department of Physics and Astronomy, University of Pennsylvania, Philadelphia, PA, 19104, USA}

\author[0000-0002-1940-4289]{Simon R. Dicker}
\affiliation{Department of Physics and Astronomy, University of Pennsylvania, Philadelphia, PA, 19104, USA}

\author{Peter N. Dow}
\affiliation{Department of Astronomy, University of Virginia, Charlottesville, VA 22904, USA}

\author[0000-0002-9693-4478]{Shannon M. Duff}
\affiliation{Quantum Sensors Division, NIST, Boulder, CO, 80305, USA}

\author[0000-0002-9962-2058]{Daniel Dutcher}
\affiliation{Department of Physics, Princeton University, Princeton, NJ, 08540, USA}

\author[0000-0001-7225-6679]{Nicholas Galitzki}
\affiliation{Department of Physics, University of Texas at Austin, Austin, TX, 78712, USA}
\affiliation{Weinberg Institute for Theoretical Physics, Texas Center for Cosmology and Astroparticle Physics, Austin, TX, 78712, USA}

\author[0000-0002-4421-0267]{Joseph E. Golec}
\affiliation{Department of Physics, University of Chicago, Chicago, IL 60637, USA}
\affiliation{Kavli Institute for Cosmological Physics, University of Chicago, Chicago, IL 60637, USA}

\author[0000-0001-9880-3634]{John C. Groh}
\affiliation{Physics Division, Lawrence Berkeley National Laboratory, Berkeley, CA 94702, USA}

\author[0000-0003-1760-0355]{Jon E. Gudmundsson}
\affiliation{Science Institute, University of Iceland, 107 Reykjavik, Iceland}
\affiliation{The Oskar Klein Centre, Department of Physics, Stockholm University, SE-106 91 Stockholm, Sweden}

\author[0000-0001-6519-502X]{Saianeesh K. Haridas}
\affiliation{Department of Physics and Astronomy, University of Pennsylvania, Philadelphia, PA, 19104, USA}

\author[0000-0002-3757-4898]{Erin Healy}
\affiliation{Kavli Institute for Cosmological Physics, University of Chicago, Chicago, IL 60637, USA}

\author[0000-0002-2781-9302]{Johannes Hubmayr}
\affiliation{Quantum Sensors Division, NIST, Boulder, CO 80305, USA}

\author[0000-0001-7466-0317]{Jeffrey Iuliano}
\affiliation{Department of Physics and Astronomy, University of Pennsylvania, Philadelphia, PA, 19104, USA}

\author[0000-0002-6898-8938]{Bradley R. Johnson}
\affiliation{Department of Astronomy, University of Virginia, Charlottesville, VA 22904, USA}

\author[0009-0000-1481-8370]{Claire S. Lessler}
\affiliation{Department of Physics, University of Chicago, Chicago, IL 60637, USA}
\affiliation{Kavli Institute for Cosmological Physics, University of Chicago, Chicago, IL 60637, USA}

\author[0009-0009-0489-8720]{Richard A. Lew}
\affiliation{Quantum Sensors Division, NIST, Boulder, CO, 80305, USA}

\author{Michael J. Link}
\affiliation{Quantum Sensors Division, NIST, Boulder, CO, 80305, USA}

\author[0000-0001-7694-1999]{Tammy J. Lucas}
\affiliation{Quantum Sensors Division, NIST, Boulder, CO, 80305, USA}

\author{Jeffrey J. McMahon}
\affiliation{Department of Astronomy and Astrophysics, University of Chicago, Chicago, IL, 60637, USA}
\affiliation{Department of Physics, University of Chicago, Chicago, IL 60637, USA}
\affiliation{Kavli Institute for Cosmological Physics, University of Chicago, Chicago, IL 60637, USA}

\author[0000-0002-7340-9291]{Jenna E. Moore}
\affiliation{School of Earth and Space Exploration, Arizona State University, Tempe, AZ, 85287, USA}

\author[0000-0002-8307-5088]{Federico Nati}
\affiliation{Departiment of Physics, University of Milano-Bicocca, Milan, MI, 20126, Italy}

\author[0000-0001-7125-3580]{Michael D. Niemack}
\affiliation{Department of Physics, Cornell University, Ithaca, NY 14853, USA}
\affiliation{Department of Astronomy, Cornell University, Ithaca, NY 14853, USA}

\author{Benjamin L. Schmitt}
\affiliation{Department of Physics and Astronomy, University of Pennsylvania, Philadelphia, PA, 19104, USA}

\author[0000-0001-7480-4341]{Max Silva-Feaver}
\affiliation{Department of Physics, Yale University, New Haven, CT 06520, USA}
\affiliation{Wright Laboratory, Yale University, New Haven, CT, 06511 USA}

\author{Robinjeet Singh}
\affiliation{Quantum Sensors Division, NIST, Boulder, CO, 80305, USA}

\author[0000-0002-1187-9781]{Rita F. Sonka}
\affiliation{Department of Physics, Princeton University, Princeton, NJ, 08540, USA}

\author[0000-0001-9528-8147]{Alex Thomas}
\affiliation{Department of Astronomy and Astrophysics, University of Chicago, Chicago, IL, 60637, USA}

\author{Robert J. Thornton}
\affiliation{Department of Physics and Astronomy, University of Pennsylvania, Philadelphia, PA, 19104, USA}

\author[0000-0002-1667-2544]{Tran Tsan}
\affiliation{Department of Physics, University of California San Diego, San Diego, CA, 92093, USA}

\author[0000-0003-2486-4025]{Joel N. Ullom}
\affiliation{Quantum Sensors Division, NIST, Boulder, CO, 80305, USA}

\author{Jeffrey L. Van Lanen}
\affiliation{Quantum Sensors Division, NIST, Boulder, CO 80305, USA}

\author[0000-0002-2105-7589]{Eve M. Vavagiakis}
\affiliation{Department of Physics, Cornell University, Ithaca, NY 14853, USA}

\author[0000-0003-2467-7801]{Michael R. Vissers}
\affiliation{Quantum Sensors Division, NIST, Boulder, CO 80305, USA}

\author[0000-0002-8710-0914]{Yuhan Wang}
\affiliation{Department of Physics, Princeton University, Princeton, NJ, 08540, USA}

\author[0000-0003-4645-7084]{Kaiwen Zheng}
\affiliation{Department of Physics, Princeton University, Princeton, NJ, 08540, USA}

\begin{abstract}
{ The Simons Observatory will map the temperature and polarization over half of the sky, at millimeter wavelengths in six spectral bands from the Atacama Desert in Chile. These data will provide new insights into the genesis, content, and history of our Universe; the astrophysics of galaxies and galaxy clusters; objects in our solar system; and time-varying astrophysical phenomena. This ambitious new instrument suite, initially comprising three 0.5~m small-aperture telescopes and one 6~m large aperture telescope, is designed using a common combination of new technologies and new implementations to realize an observatory significantly more capable than the previous generation. In this paper, we present the pre-deployment performance of the first mid-frequency ``optics tube'' which will be fielded on the large aperture telescope with sensitivity to the 90 and 150~GHz spectral bands. This optics tube contains lenses, filters, detectors, and readout components, all of which operate at cryogenic temperatures. It is one of seven that form the core of the large aperture telescope receiver in its initial deployment. We describe this optics tube, including details of comprehensive testing methods, new techniques for beam and passband characterization, and its measured performance. The performance metrics include beams, optical efficiency, passbands, and forecasts for the on-sky performance of the system. We forecast a sensitivity that exceeds the requirements of the large aperture telescope with greater than 30\% margin in each spectral band, and predict that the instrument will realize diffraction-limited performance and the expected detector passbands.}
\end{abstract}

\section{Introduction}
Measurements of the Cosmic Microwave Background (CMB) underlie our modern understanding of the Universe from its earliest moments~\citep{Planck2018,Choi_2020,Dutcher_2021,bicep2021}, through the formation and evolution of large-scale structure~\citep{PlanckLensing,qu2023atacama,Millea_2021}, and up to the present~\citep{Hensley_2022,Naess_2021}. Improving these measurements is critical to addressing fundamental questions, including the physics of the primordial universe~\citep{Abazajian_2015,linde2005particle,PlanckInflation2016}, the possibility of new light relativistic particles~\citep{Bashinsky_2004}, and the measurement of the neutrino mass sum~\citep{Couchot_2017}, as well as many additional astrophysical and cosmological questions~\citep{Ade_2019,abazajian2016cmbs4}.

This abundance of scientific opportunity has motivated a series of CMB experiments both on the ground~\citep{Benson_2014,Niemack_2010,Essinger_Hileman_2014,Ahmed_2014,Inoue_2016} and in space~\citep{Bennett_1996,Bennett_2003,PlanckOverview} to image the millimeter-wave sky. The ground-based efforts include the use of large telescopes and offer the ability to integrate deeply on significant fractions of the sky~\citep{Essinger_Hileman_2014,Niemack_2010,Ade_2019}. Both the Atacama Cosmology Telescope (ACT) and South Pole Telescope (SPT) teams have fielded three generations of cameras with increasing capabilities on large telescopes with arcminute scale resolution~\citep{Carlstrom_2011,Swetz_2011,Thornton_2016,Austermann_2012,Benson_2014,Henderson_2016}. The scientific results from these experiments are nearing the constraining power of \emph{Planck} and are expected to improve with data already on hand. Pushing to the sensitivities required to realize the full potential of the millimeter-wave sky requires significantly more capable telescopes and instruments. Since the detectors have already reached the fundamental noise limits~\citep{Staggs_review}, large gains in sensitivity can only be achieved by fielding arrays with more detectors in instruments with expanded fields of view. Two such designs have been developed~\citep{Niemack_2016,gallardo2022optical} for the next generations of telescopes, including Simons Observatory~\citep{Ade_2019} and CMB-S4~\citep{abazajian2016cmbs4}. 

The Simons Observatory is a suite of millimeter-wave telescopes located at an altitude of 5200~m on Cerro Toco in the Chilean region of the Atacama Desert. In the nominal design, it includes a large aperture telescope (6~m LAT) as well as three small-aperture telescopes (0.5~m SATs). Together, these instruments will cover a broad frequency range spanning 30 to 280~GHz at both large and small angular scales. With an initial configuration of over 60,000 detectors, Simons Observatory will significantly advance the state of cosmology.

This paper describes an effort to validate the performance of the large aperture telescope cryogenic optics and detectors prior to deployment to Chile. We present characterization of a large aperture telescope mid-frequency optics tube, operating with detectors at spectral bands centered near 90 and 150~GHz, around the CMB peak frequency. Section~\ref{sec:science} motivates the need for an exhaustive testing program that targets the largest performance drivers of the camera. Section~\ref{sec:testing} gives an overview of the large aperture telescope optics tube in-lab testing program. Section~\ref{sec:bandpass} describes the passband characterization of all 90 and 150~GHz channels across the focal plane using a combination of methods. Section~\ref{sec:beams} covers the characterization of the instrument beam using both a coherent radio source and a non-coherent thermal source. Section~\ref{sec:polarization} describes the polarization performance through the optics tube. Sections ~\ref{sec:det_perform} and \ref{sec:detector-gain} detail the extensive detector characterization performed in an integrated optical system. Section~\ref{sec:efficiency} describes the measured end-to-end efficiency of the detectors and optics. Section~\ref{sec:on-sky} details the predicted on-sky performance using the results of this testing program.

\section{Instrument overview and performance drivers} \label{sec:science}
\begin{figure*}[t]
    \centering
    \includegraphics[width= \linewidth]{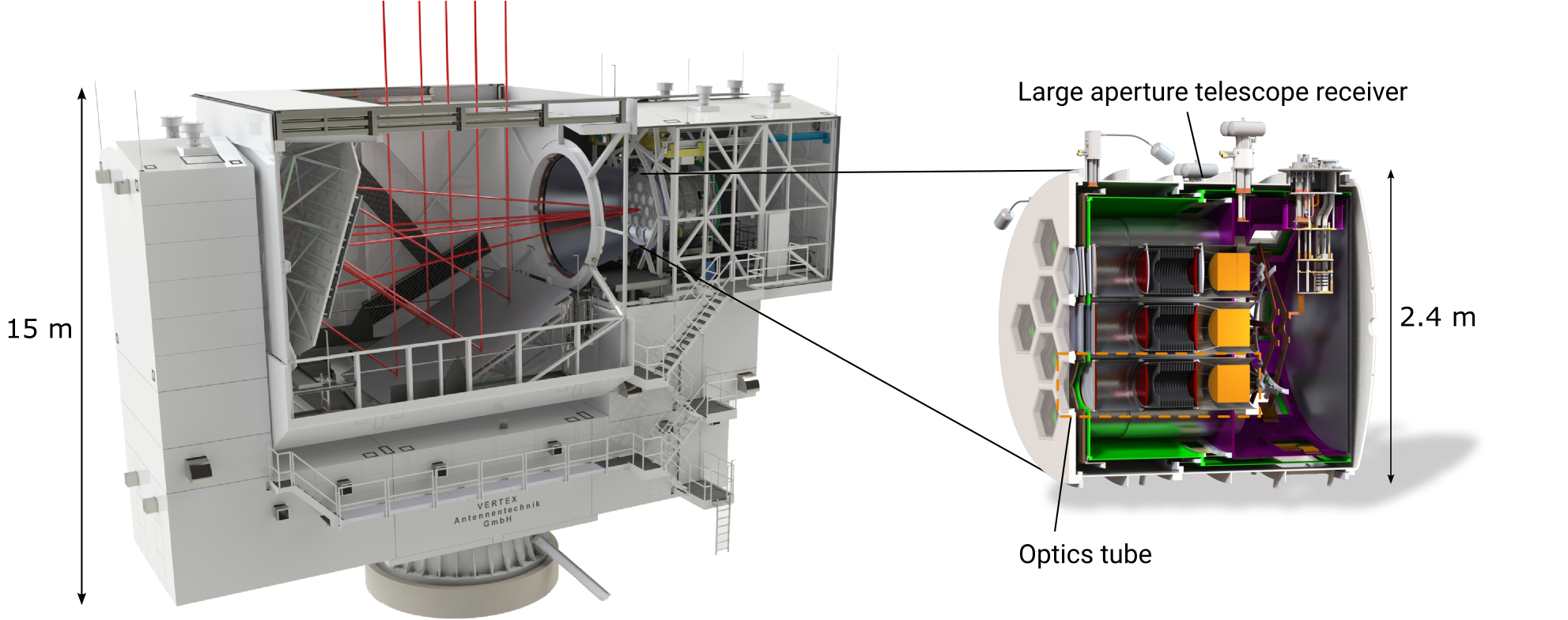}
    \caption{\textit{Left}: The Simons Observatory large aperture telescope is a 6~m crossed-Dragonne telescope with a large field-of-view designed to illuminate over 60,000 detectors. \textit{Right}: The large aperture telescope receiver (LATR) couples light from the sky via the telescope's two primary mirrors. When fully populated, the receiver holds 13 optics tubes. Each optics tube is sensitive to one of three dichroic passbands: 30 / 40 GHz (low-frequency), 90 / 150~GHz (mid-frequency), and 220 / 280~GHz (ultra-high frequency). The goal of the LATR-Tester program is to fully test and validate each type of optics tube prior to deployment on the large aperture telescope. Details of the optics tube are shown in Figure \ref{fig:latrt}.}
    \label{fig:LAT}
\end{figure*}

The 6~m Simons Observatory large aperture telescope is a new crossed-Dragonne telescope with a diffraction-limited field-of-view of $7.8\degree$ at 90~GHz. The cryogenic receiver for the large aperture telescope~\cite{Zhu2021_LATR} is 2.4~m in diameter and couples light from the telescope's two mirrors onto the detectors. The receiver will be initially deployed with 30,000 detectors, but will be increased to 60,000 detectors for the future Advanced Simons Observatory upgrade. The large aperture telescope and its receiver are shown in Figure \ref{fig:LAT}.

Within the large aperture telescope receiver, the detectors are portioned into units we refer to as optics tubes~\citep{Kofman_2024}, which are composed of anti-reflection coated lenses~\citep{Golec_2020}, filters~\citep{Golec_2022,Cardiff_2006}, internal baffling~\citep{Xu_2021}, detector arrays~\citep{McCarrick2021_MFUFM}, and a microwave multiplexing detector readout~\citep{dober2020microwave}. Each optics tube is designed to be fielded with one type of dichroic detector array spanning an octave bandwidth. The three array types are low-frequency (30 / 40~GHz), mid-frequency (90 / 150~GHz), and ultra-high-frequency (220 / 280~GHz). Simons Observatory will initially field the large aperture telescope with seven optics tubes: one at low-frequency, four at mid-frequency, and two at ultra-high-frequency.

The science goals of the large aperture telescope place well-defined requirements on the performance of the instrument \citep{Ade_2019}. In addition to white noise performance, instrumental systematics must also be well understood. The requirement for passband uncertainty is driven by component separation, which requires the uncertainty in relative gain and center frequency of each array-averaged spectral band to be less than 2\% and 1\%, respectively~\citep{Ward_2018}. For small angular scale science targets, like the number of relativistic particles $N_\textrm{eff}$ and the neutrino mass sum, the beam shape must be calibrated to a similar precision~\citep{Wallisch_2018}. Meeting these requirements is a careful process of constraining white noise levels, and predicting and controlling potential systematic effects.

These factors motivate the need for an instrument model that can identify key areas where performance can have a significant impact on the measurement goals. In terms of raw noise performance, the sensitivity of the optics tube is determined by a number of factors including: the optical efficiency of the detectors, the total number of working detector channels (yield), signal losses due to imperfections in the optics, spurious couplings to 300~K signals from the telescope and surroundings, and deviations from the designed passband which could lead to reduced bandwidth or unexpected coupling to atmospheric emission. These metrics all have direct and well-understood impacts on the projected sensitivity of the instrument and are considered to be the key performance drivers to be tested prior to deployment. For example, if the fraction of the beam that spills over to 300~K is 6\% instead of the desired 1\%, that would represent a nearly 50\% mapping speed loss in the 90 and 150~GHz bands. Similar assessments can be made for unwanted power coupling to emission lines from the atmosphere. Losses in optical efficiency, such as from low detector coupling, reflective filters, and cross-polarization, will translate linearly to losses in mapping speed, which grows as the inverse-square of instrument noise. Some of these quantities, such as detector efficiency, can be verified with subsystem tests. But many other tests, such as passbands and leakage from 300~K sources, depend on the full optical coupling system that includes the filters, detectors, and optical elements, and can only be quantified with integrated tests of the full optics tube.

A number of instrumental systematic effects must also be controlled to meet our science goals. These include effects such as optical time constants, scattering and diffractive sidelobes, passband mismatch across the detector array, long-term stability of optically-loaded detectors, and the overall beam quality of the instrument. These can only be tested in a fully-integrated optical system.

\section{\label{sec:testing}Testing Program}
\begin{figure*}[t]
    \centering
    \includegraphics[width = \textwidth]{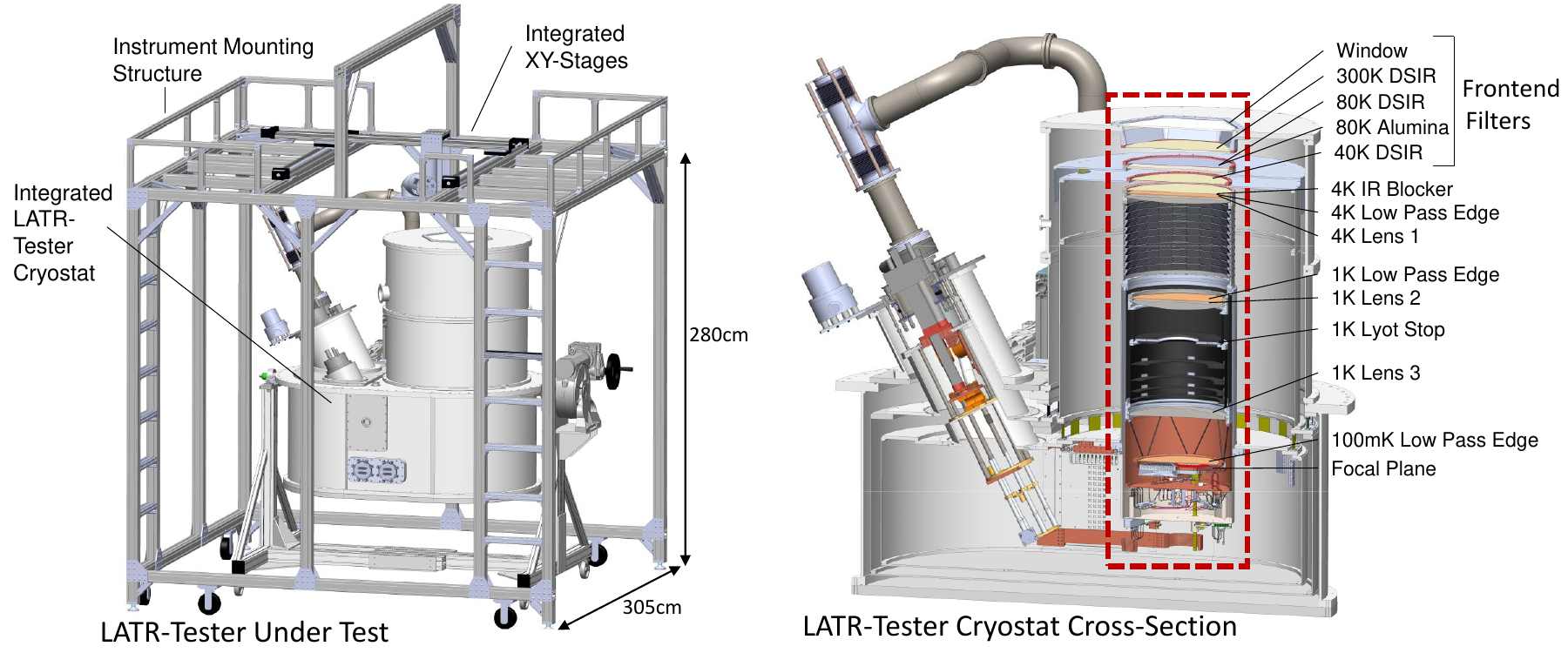}
    \caption{\textit{Left}: The Simons Observatory Large Aperture Telescope Receiver Tester (LATR-Tester or LATRt) cryostat is shown in context with the Instrument Mounting Structure. The mounting structure has integrated XY-stages that were used to move the different optical test equipment described throughout this paper. The structure was designed so it could be easily rolled over the LATRt cryostat after test equipment is installed, significantly simplifying the process of changing between types of measurements. \textit{Right}: The LATR-Tester cross-section with a single integrated optics tube and associated frontend filtering, indicated by the dashed box. Light enters through a high-density polyethylene window, travels through a series of double-side infrared-blocking filters (DSIR), including one alumina IR-blocking filter, and passes through the 4~K low-pass-edge (LPE) mesh filter. The light is focused on the detector arrays by the three lenses, with two additional low-pass-edge filters near the 1~K and 100~mK stages. The optics tube components are further described in~\cite{Kofman_2024} and \cite{Zhu2021_LATR}.}
    \label{fig:latrt}
\end{figure*}

The goal for this work is to validate the sensitivity and control of systematics of the Simons Observatory large aperture telescope optics tubes prior to deployment. The comprehensive suite of tests necessary for this validation calls for a dedicated laboratory testing setup with shorter turnaround times and greater flexibility compared to the full-size large aperture telescope receiver. This program requires the integration of complete assembled optics tubes in a cryogenic receiver with full detector readout data acquisition systems, data analysis, and a variety of test equipment. The fully-integrated system is used to perform a variety of optical tests, including: efficiency with both internal cryogenic and external thermal loads; beam maps with a coherent holography system and with thermal sources; passband measurements with a Fourier-transform spectrometer (FTS), laser-derived coherent source, and thick grill filters; time constants as measured using a chopped source; and a variety of detector tests. When completed, this program will have fully characterized one large aperture telescope optics tube of each dichroic band set: low-frequency, mid-frequency, and ultra-high-frequency. This paper describes the test system, measurements, and results for the tests performed on a mid-frequency optics tube.

\subsection{\label{sec:cryostat}Cryostat}

The test cryostat~\citep{Harrington2020} was adapted from a Simons Observatory small-aperture telescope~\citep{Ali_2020} to accommodate the installation of a large aperture telescope optics tube with nearly identical optical and thermal interfaces as in the large aperture telescope receiver (LATR) and with a significantly faster cooldown time to efficiently facilitate multiple rounds of lab testing. This cryostat is referred to as the LATR-Tester and is shown in Figure~\ref{fig:latrt}.  

The key changes from the small-aperture telescope design include the removal of the cryogenic half-wave plate interface, the addition of an 80~K mounting stage, the use of a large aperture telescope receiver vacuum window, and the use of a filtering scheme that is equivalent to that of the large aperture telescope receiver~\citep{Zhu2021_LATR}. The vacuum shell and 40~K shell were lengthened to provide room for the large aperture telescope optics tube and an adapter plate was added to mount the optics tube. Since the small-aperture telescope lacks the 80~K cooling stage of the large aperture telescope receiver, the 80~K filters were mounted in their nominal optical position using an adapter built off the 40~K stage.

 \subsection{\label{sec:readout}Optics Tube}
The optics tube~\citep{Kofman_2024} contains cryogenic lenses, filters, a Lyot stop, and a focal plane containing three detector arrays. These components are heat-sunk at various cryogenic temperatures at 4~K or colder, and is surrounded by baffles as shown in Figure~\ref{fig:latrt}. In a time-reversed analysis of this optical system, light emitted from the detector diffracts through a spline-profile feedhorn \citep{Simon_2018}, which defines the detector beams. The next elements in the optical path are a low-pass filter at 100~mK, and a lens and series of blackened ring baffles at 1~K. At 1.5~K, there is a Lyot stop, second lens and second low-pass filter. At 4~K, there is a section of metamaterial microwave absorbing tiles~\citep{Xu_2021} that are designed to control stray light prior to the colder stages of the optics tube. Additionally at 4~K, there is a third lens and low-pass filter, as well as an infrared-blocking filter. The three low-pass filters \citep{Cardiff_2006} are designed to block out-of-band radiation, with cutoffs close to but not defining the passbands of the instrument, which are instead defined by the on-chip filters and feedhorn waveguide on the detector array. 

Outside of the optics tube, there are the additional thermal filters mentioned in the previous section, including infrared-blocking filters at 40~K, 80~K, and 300~K and an anti-reflection coated alumina filter at 80~K~\citep{Golec_2022}. In addition to rejecting thermal radiation, the alumina filter doubles as a prism to center the beams on the telescope for those optics tubes that are placed off-center in the large aperture telescope receiver. The final component in this chain is a 3.1~mm thick anti-reflection coated window made of high-density polyethylene.

The cold optics design was chosen to limit the amount of thermal and optical loading on the detectors while re-imaging the detector array onto the telescope focal plane, correcting aberrations for good image quality, and providing a Lyot stop to control stray light. Our testing program is designed to verify the performance of these components while also checking for performance degradation from effects like diffraction, ghosting, and unexpected scattering.

The detector arrays are optimized for the on-sky optical powers expected in Chile and will saturate when exposed to typical 300~K lab conditions. To enable laboratory testing with these detectors, we install neutral density filters (NDF) to strongly attenuate the incoming signal. We add mounts (used only in testing) to hold the NDFs between the focal plane and the first 100~mK low pass edge filter, or between the 80~K alumina filter and the 40~K double sided IR-blocking filter. After initial tests, we found the first position closest to the focal plane to be best for limiting ghosting effects in the beam.

\subsection{\label{sec:fpb} Focal-plane Setup}
Each optics tube has three detector array positions available at the focal plane. This flexibility allows us to incorporate different types of detectors to optimize testing. These include a standard Simons Observatory focal-plane module, which houses the detector array; a single-pixel box with high thermal conductance to the bath; and a holography receiver with a mixer coupled to a feed horn.

The focal-plane module~\citep{McCarrick_2021} consists of about 1800 transition-edge sensor (TES) bolometers, which are read out with two microwave-multiplexing~\citep{dober2020microwave} amplification chains. Each detector pixel is dichroic and uses feedhorn waveguides coupled to planar orthomode transducers to split incoming light into orthogonal polarizations~\citep{Datta_2014}. The detector array used in this testing was a prototype, which affects certain thermal and mechanical properties but is not expected to impact optical performance. In addition to the standard optical components, an NDF is placed in the optical path of this detector module in order to achieve the expected loading during observation.

The single-pixel box consists of a single feedhorn coupled to several ``high-$G$'' bolometers, where $G$ refers to the thermal conductance between the bolometer absorber and bath. The higher thermal coupling increases the saturation powers in these detectors, allowing for direct observations of a 300~K laboratory environment without saturating and without the use of an NDF.

\begin{figure}[ht]
    \centering
    \includegraphics[width=0.45\textwidth]{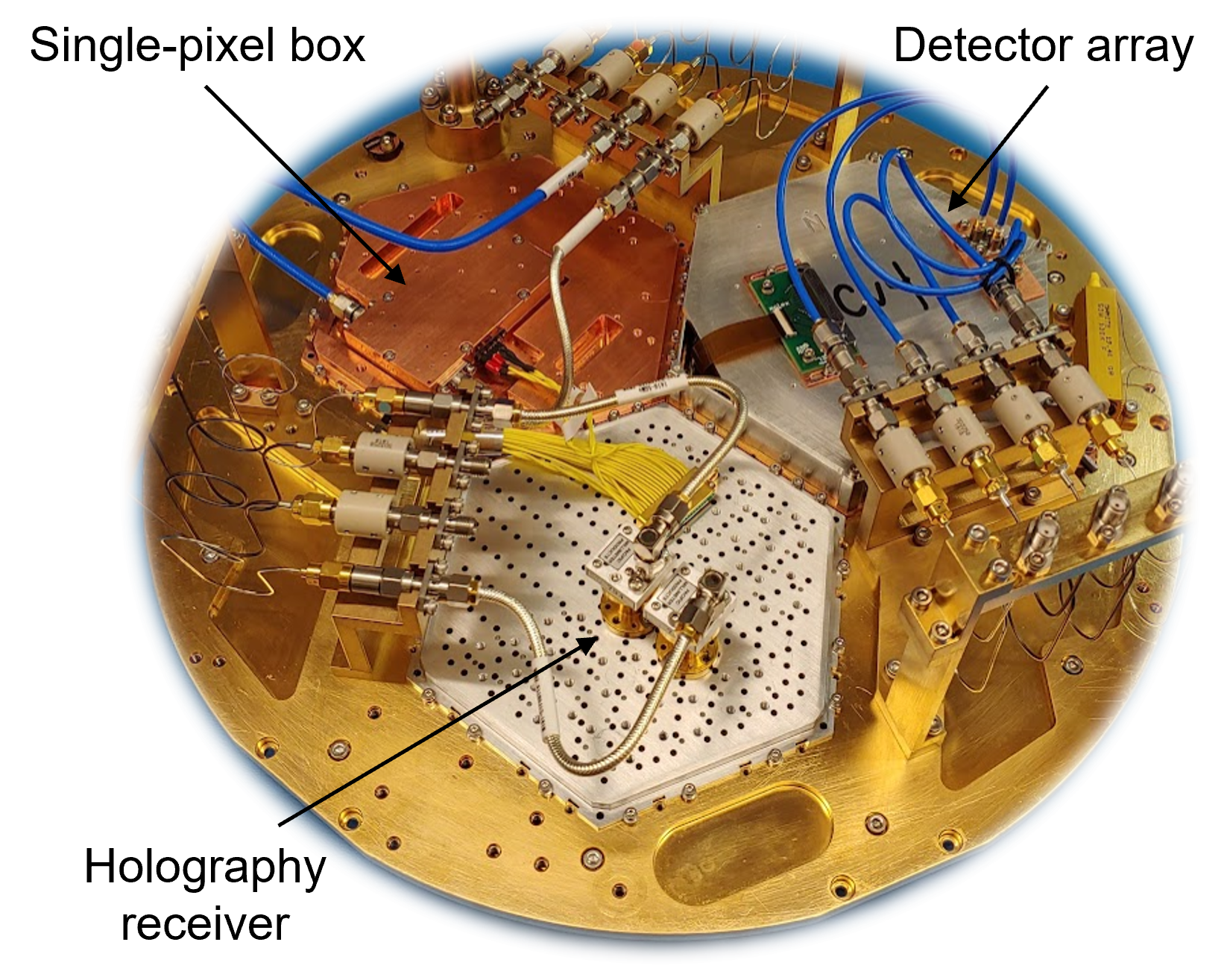}
    \caption{The 100~mK focal plane configuration inside the LATR-Tester optics tube (back view). In this testing configuration, the three nominal detector array positions of the optics tube focal plane are occupied by one detector array, a high-$G$ single-pixel box, and a radio holography receiver.}
    \label{fig:focal-plane}
\end{figure}

The setup for radio holography consists of a Simons Observatory feedhorn array modified to include connections for standard waveguide flanges at the outputs, to which a round to rectangular waveguide transition and a harmonic mixer capable of operating at 4~K are mounted. As described in Section~\ref{sect:holog-method}, this setup allows for a complete holographic measurement of the optics tube beam. As is the case with the single-pixel box, no NDF is necessary to operate the holography receiver.

The three devices installed in the focal plane, the detector array, the high-$G$ single pixel, and holography array, remained the same throughout the entirety of this optics tube testing.

\subsection{\label{sec:fts_ndf}Neutral Density Filter Calibration}

\begin{figure}
    \centering
    \begin{tabular}{cc}
    \includegraphics[width=0.21\textwidth]{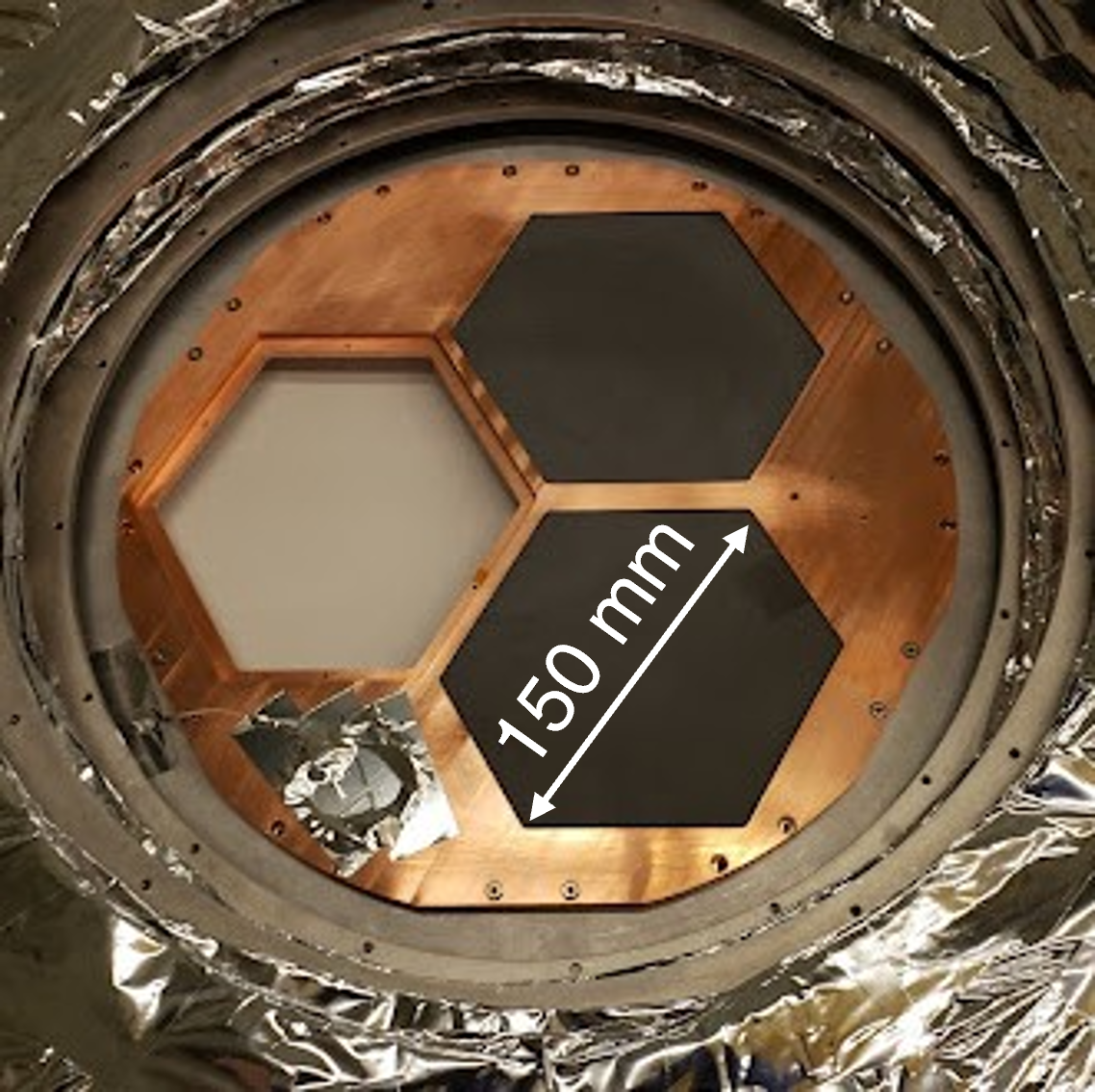} &  
    \includegraphics[width=0.24\textwidth]{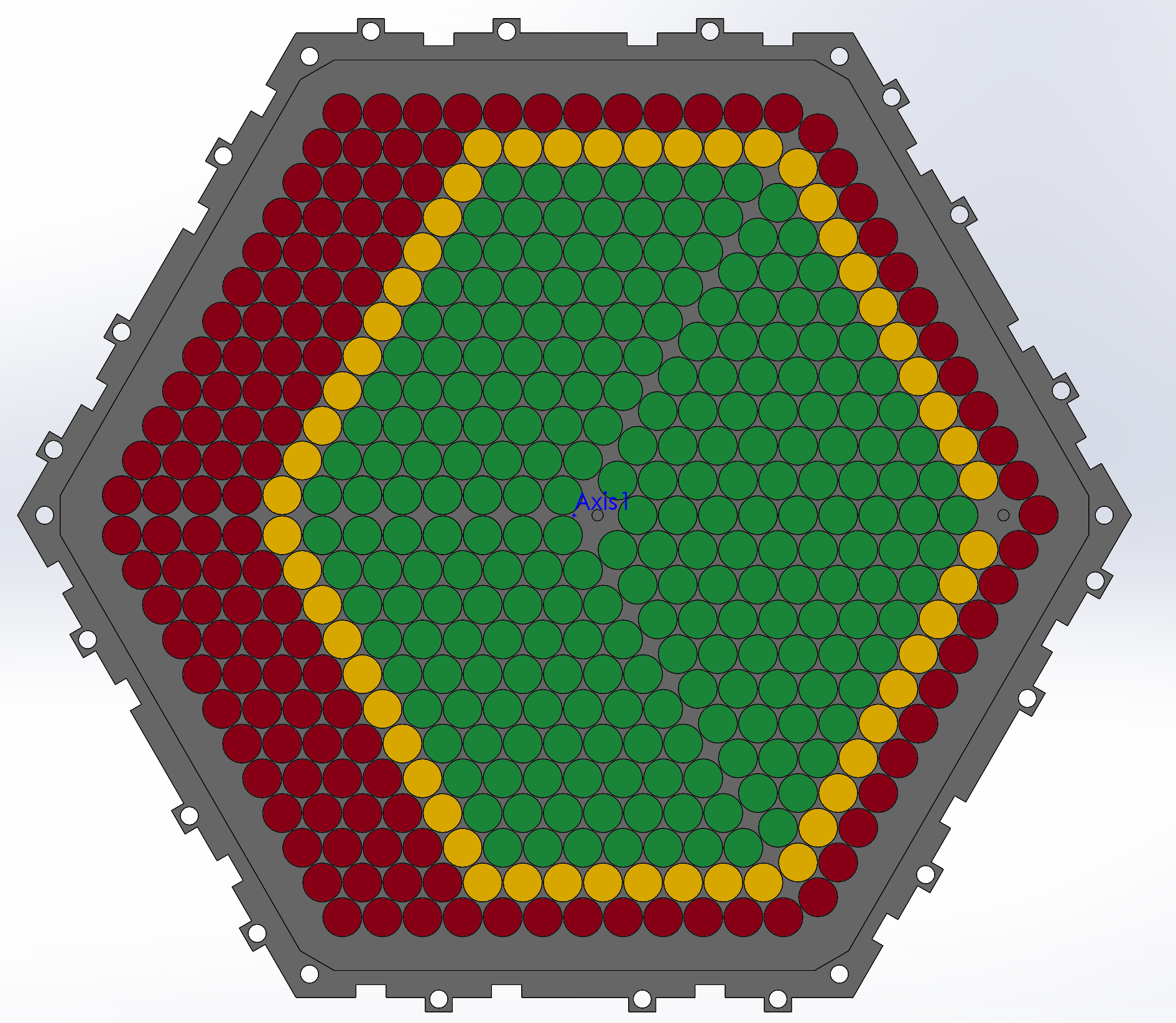} \\
    \end{tabular}
    \caption{\label{fig:ndf_mounting} \emph{Left}: The segmented NDF mounting plate shown mounted at the 4~K end of the optics tube with two NDF filters installed. \emph{Right}: A color coding of the pixels of one detector array indicating which pixels are expected to be unobstructed (green), partially obstructed (yellow), and fully obstructed (red) by the segmentation of the NDF mounting plate at a mounting position of 20~mm from the focal plane at the sky side of the 100~mK low-pass edge filter.}
\end{figure}

The detectors in the optics tube are designed to operate while observing through the Chilean atmosphere, where the typical optical loading is expected to be 1.0 and 2.3~pW for the 90~and 150~GHz detectors. Conversely, a 300~K laboratory will produce 21 and 41~pW of loading in the 90~and 150~GHz bands, meaning that the detectors will saturate without additional filtering. In the testing described here, we use an absorptive NDF that prevents at least $\sim95\%$ of the incident in-band power from reaching the detectors. Despite its name, the transmission of the NDF used in the LATR-Tester is not constant across the measurement frequencies of interest. Accurately calibrating both the absolute and relative transmission of the NDF at the temperatures of operation in these tests is critical for receiver characterization.

Eccosorb~MF\footnote{https://www.laird.com/products/microwave-absorbers/injection-molded-machined-cast-liquids-and-microwave-absorbing-thermoplastic/eccosorb-mf}, a rigid machinable absorptive material, was chosen as the NDF material because pre-prepared 1~ft by 1~ft sheets can be obtained in a variety of thicknesses and absorptive strengths. In particular, we selected Eccosorb~MF114 because a 6.35~mm thick sheet provided sufficient loss in the 90~GHz band and was thin enough to fit in the space available in our optics tube. Due to concerns over delamination, the NDF filters were not anti-reflection coated.

The left image in Figure~\ref{fig:ndf_mounting} shows the segmented NDF mounting plate which is designed to hold up to three NDFs and can be mounted at focal points on the sky side of the 100~mK low-pass-edge filter or at the 4~K end of the optics tube. A segmented mount was used because of the size limitations of the NDF materials and the desire to be able to run both the high-$G$ single pixel and holography arrays without NDFs. The NDF mount was placed at the end of the optics tube during the first two optical cooldowns, Cooldown 1 and Cooldown 2, because this placement made it significantly easier to change NDF configurations between cooldowns. An NDF was installed in front of the high-$G$ single pixel in Cooldown 1 and then removed for Cooldown 2. Relative measurements between the two cooldowns were then used to calibrate the NDF transmission.

The calibration plan for the absolute NDF transmission as a function of frequency initially included both FTS measurements, described in Section~\ref{sec:bandpass} and end-to-end responsivity measurements, described in Section~\ref{sec:efficiency}, of the high-$G$ single pixel box with and without the NDF. However, low optical efficiency and higher instabilities of the high-$G$ single-pixel channels under low optical loading conditions invalidated the end-to-end responsivity measurements. For this reason, coherent reflection and transmission measurements at 300~K were used to constrain the model where necessary.

Following~\cite{Halpern_1986}, the transmission function of the NDF is modeled as
\begin{equation}
    \label{eq:ndf}
    T(\nu) = T_{0} \exp(-a\nu^b)
\end{equation}
\noindent where the zero-frequency transmission, $T_0 = 0.82$, was determined through coherent reflection measurements by fitting the asymptotic reflectance at high frequency: $T_0 = 1-R(\nu)|_{\nu \rightarrow \infty}$. The spectral index, $b = 1.38 \pm 0.04$, was taken from fits to the 300~K transmission measurements over the frequency range of interest. Although this parameter is expected to change slightly at cryogenic temperatures~\citep{Halpern_1986}, this change is expected to be subdominant to our measurement uncertainties.

\begin{figure*}[t]
    \centering
    \includegraphics[width=0.9\textwidth]{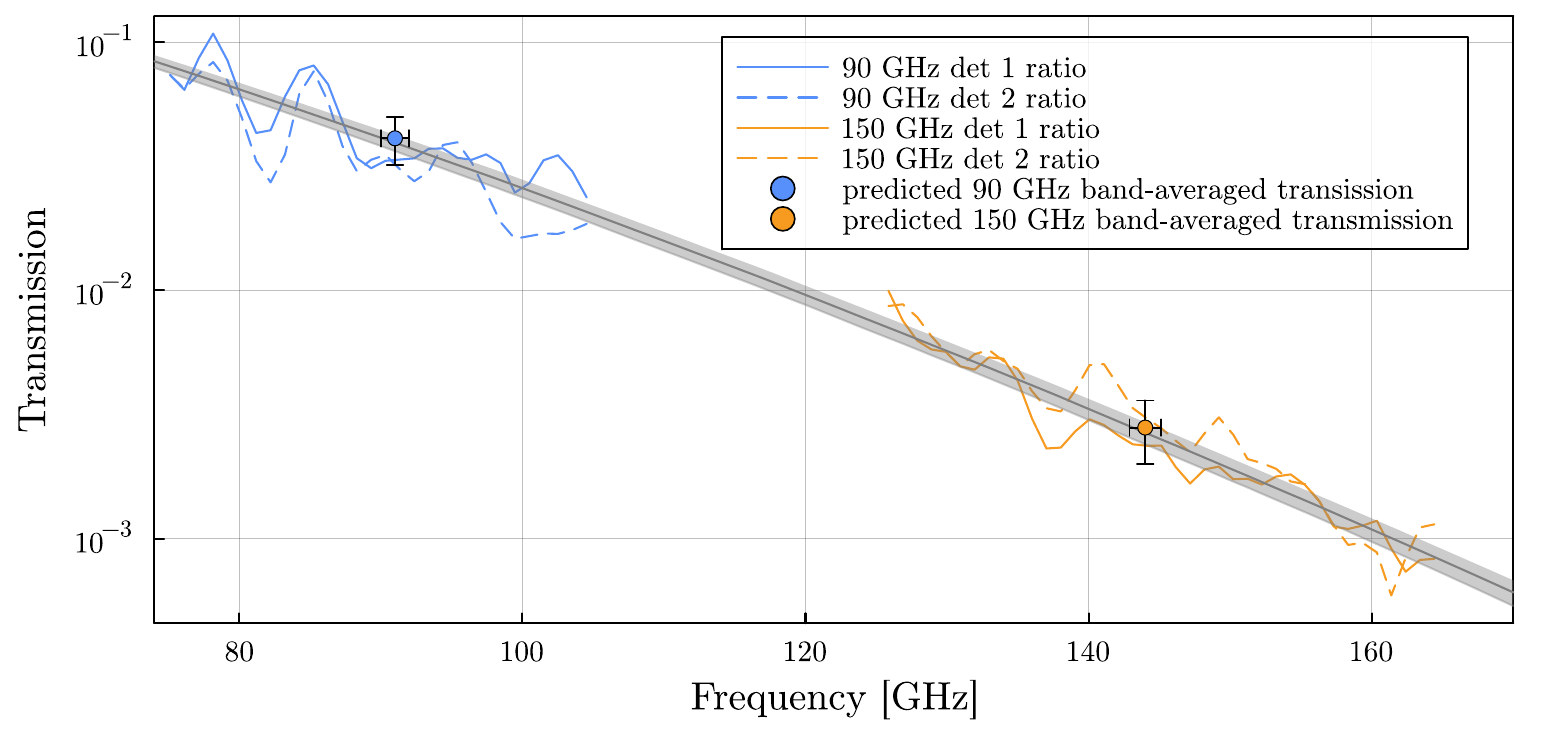}
    \caption{Single pixel measurements used for NDF calibration, consisting of two 90~GHz and two 150 GHz detectors. Each curve shows the ratio of measurements taken with and without the NDF in front of the detectors. The NDF frequency dependence was fitted after combining all four detector ratio data sets and is plotted along 200 propagated samples which represent the error bars on $a$ and $b$ in Equation \ref{eq:ndf}. Oscillations in the ratio curves are likely due to swapping from an alumina filter without an anti-reflection coating in Cooldown 1 to a filter with an anti-reflection coating in Cooldown 2.}
    \label{fig:SPB_fits}
\end{figure*}

The absorption coefficient, $a$, was found using FTS measurements of the high-$G$ detectors with and without the NDF. The ratio of spectra measurements with and without the NDF, shown in Figure~\ref{fig:SPB_fits} were fit to the exponential decay function of the NDF model. The final fitted values for $a$ were $0.0059 \pm 0.0015$. Our final fitted transmission function yields band-averaged NDF transmission values of $4.2 \pm 1.0 \%$ at 90 GHz and $0.30 \pm 0.12 \%$ at 150 GHz. Because the relative NDF transmission function within the band regions is degenerate between parameter choices of $a$ and $b$, evaluating the transmission uncertainty requires careful propagation of the uncertainties and covariances of these two fit parameters.

The second optical cooldown, Cooldown 2, was originally planned to be the final optical characterization configuration for testing the mid-frequency optics tube. However, as described later in Section \ref{sec:beams}, the NDF plate mounted at 4~K end of the optics tube was found to cause reflections and spill around the open positions in the plate, significantly affecting the quality of the measurements when the test equipment did not include focusing optics. Therefore, once the NDF calibration was complete, the NDF plate was moved to the 100~mK stage where it is about about 30~mm from the focal plane. The right image in Figure~\ref{fig:ndf_mounting} indicates that about 25\% of the pixels in the detector array are expected to be partially blocked by the segmented mounting when placed on the 100~mK stage, these pixels are cut from results reported here. 

\subsection{\label{sec:readout}Detector Readout}
The detectors are read out with a microwave-multiplexing scheme, in which TES bolometers are inductively coupled via superconducting quantum interference devices (SQUID) to microwave resonators fabricated on multiplexer chips~\citep{dober2020microwave}. The response of each detector modulates the frequency of one microwave resonator, whose phase is tracked with SLAC Superconducting Microresonator Radio Frequency (SMuRF) electronics~\citep{Yu2022_SMURF}. In addition to tracking the resonators, the warm readout electronics provides the bias power for the detectors, the flux-ramp modulation to linearize the SQUID response, and the bias power for the two cryogenic amplifiers at 4\,K and 40\,K. The readout chain is completed with coaxial cables and additional passive components, described in detail in~\cite{Rao20}. This readout scheme enables up to 910 SQUID channels to be read out on one chain, the highest multiplexing achieved in a CMB instrument~\citep{McCarrick_2021}. Each detector module contains two of these multiplexing units. 

A key component of the readout chain common to both the large- and small-aperture telescopes is the Universal Readout Harness~\citep{moore2022_URH}, which houses the readout components traversing 300~K to 4~K and the 40\,K amplifiers for up to 24 readout chains. Below 4~K, the readout components, including the 4~K amplifiers, are built into the optics tube. The only readout components that are specific to the LATR-Tester are isothermal cables connecting the Universal Readout Harness to the optics tube, which emulate the coaxial ``highways'' in the large aperture telescope readout design~\citep{Zhu2021_LATR}. Similarly, extending the small-aperture telescope cryogenic thermometry to the LATR-Tester setup required custom lengths of existing cable designs, a custom 4~K breakout cable, and a custom warm breakout board to split 100\,mK signals from the warmer stages. All other components are part of the standard housekeeping plans for both the large- and small-aperture telescopes.

\subsection{Instrument Mounting Structure and Test Equipment}

\begin{figure}[ht]
    \centering
    \includegraphics[width=0.47\textwidth]{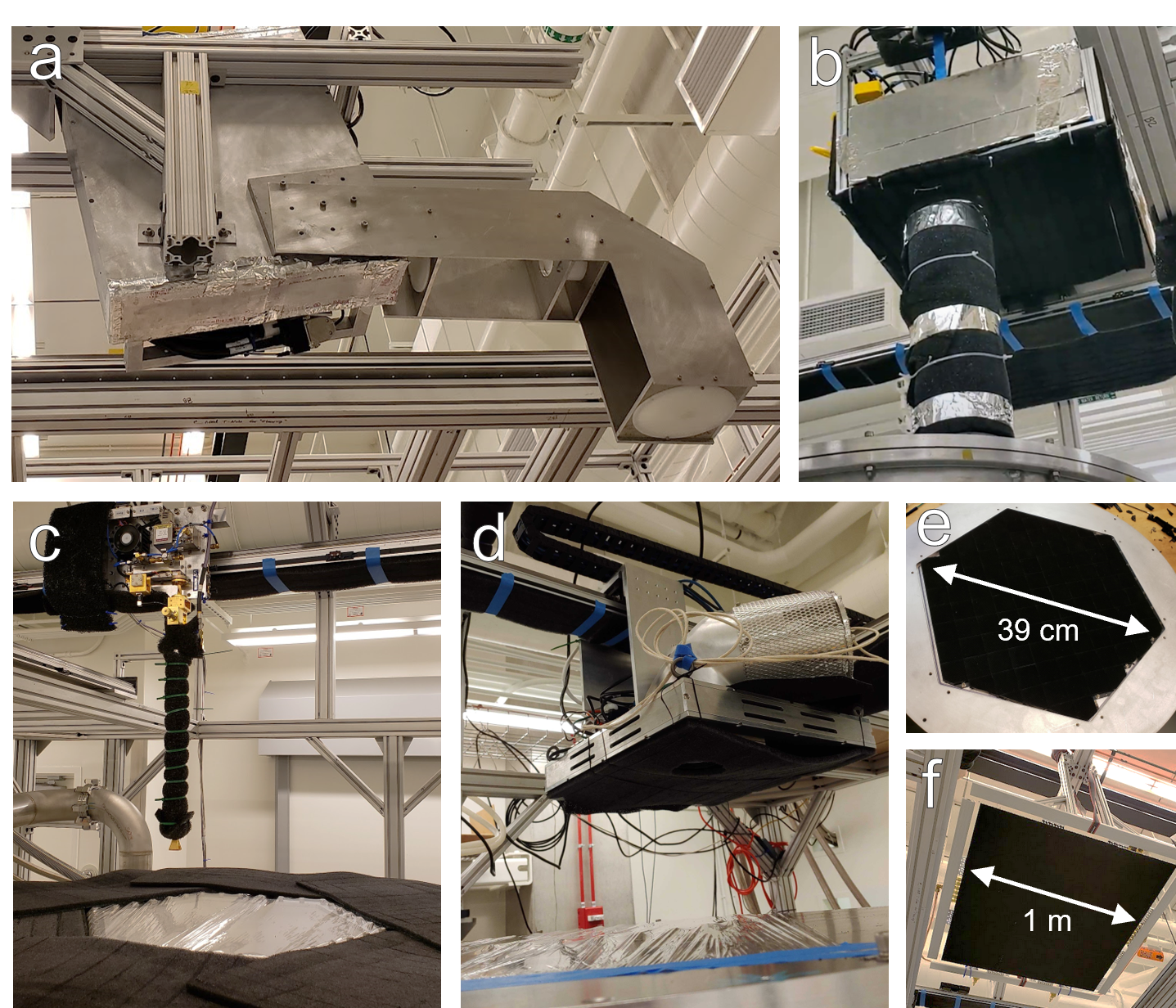}
    \caption{Test equipment that was built and used for testing on the LATR-Tester includes a) Fourier-transform spectrometer (FTS), b) frequency-tunable laser source (FLS), c) radio holography source, d) thermal chopped source, e) internal beam-filling cold load, and f) external beam-filling warm load. Equipment a) – d) mount to XY-stages on the surrounding instrument mounting structure, allowing them to be scanned across the optics tube's field of view.}
    \label{fig:equipment}
\end{figure}

A full optical characterization of the instrument requires a wide array of testing hardware. The full suite of test equipment built for the LATR-Tester program is shown in Figure \ref{fig:equipment} and will be discuss in subsequent sections. This includes: 
\begin{itemize}
    \item Fourier-transform spectrometer (FTS) \citep{Alford_2024} to measure passbands and out-of-band leaks at lower frequencies
    \item Refractive coupling optics to reimage the output of the FTS into a focused beam at the focal plane
    \item Frequency-tunable coherent laser source (FLS) \citep{fls2024} to verify band edges and probe out-of-band leaks at high frequencies
    \item Coherent radio holography \citep{chesmore2022} source/receiver pair used to fully characterize the beam out of the optics tube
    \item Thermal chopped source \citep{Sierra_thesis} to measure detector beams, time constants, and constrain out-of-band leaks at high frequencies when coupled with a high-pass filter
    \item Internal cold load (4~K) and external warm load (300~K) to measure optical efficiencies in a dark and optical configuration, respectively
\end{itemize}
To mount this equipment in a testing position, an instrument mounting structure surrounds the LATR-Tester as shown in Figure \ref{fig:latrt}. All equipment (except for the beam-filling loads) are mounted onto XY-stages that are integrated into the mounting structure, allowing the test equipment to scan back and forth in any desired pattern across the field-of-view of the optics tube. This is particularly useful for obtaining wide beam maps and probing detector properties across many positions. The design of the instrument mounting structure allows it to be easily rolled in and out of position between different tests.

 \subsection{\label{sec:cooldowns}Cooldowns}
A total of five cooldowns were completed with this cryostat using the mid-frequency optics tube. The first, designated Cooldown 0, was a dark cooldown where the frontend optics were replaced with blanks and a cold load was installed at the 4~K end of the optics tube. This dark cooldown was used to measure the detector efficiencies and dark noise properties, described in Sections~\ref{sec:cold_load} and~\ref{sec:det_perform} respectively. The first two optical cooldowns, Cooldown 1 and Cooldown 2, were used to calibrate the NDF transmission described in Section~\ref{sec:fts_ndf}. 

In the third optical cooldown, Cooldown 3, the nominal optics stack for the mid-frequency optics tube was fully characterized. The holographic beam maps, described in Section~\ref{sect:holog-method}, showed a frequency dependent spill which was attributed to the low-pass edge filter mounted near the 1~K lens. In the last cooldown, Cooldown 4, this filter was removed, and the system was fully characterized again. Unless otherwise indicated, the results presented in this paper are from Cooldown 4.

\section{\label{sec:bandpass}Spectral Response}
\begin{figure*}
    \centering
    \includegraphics[width=\linewidth]{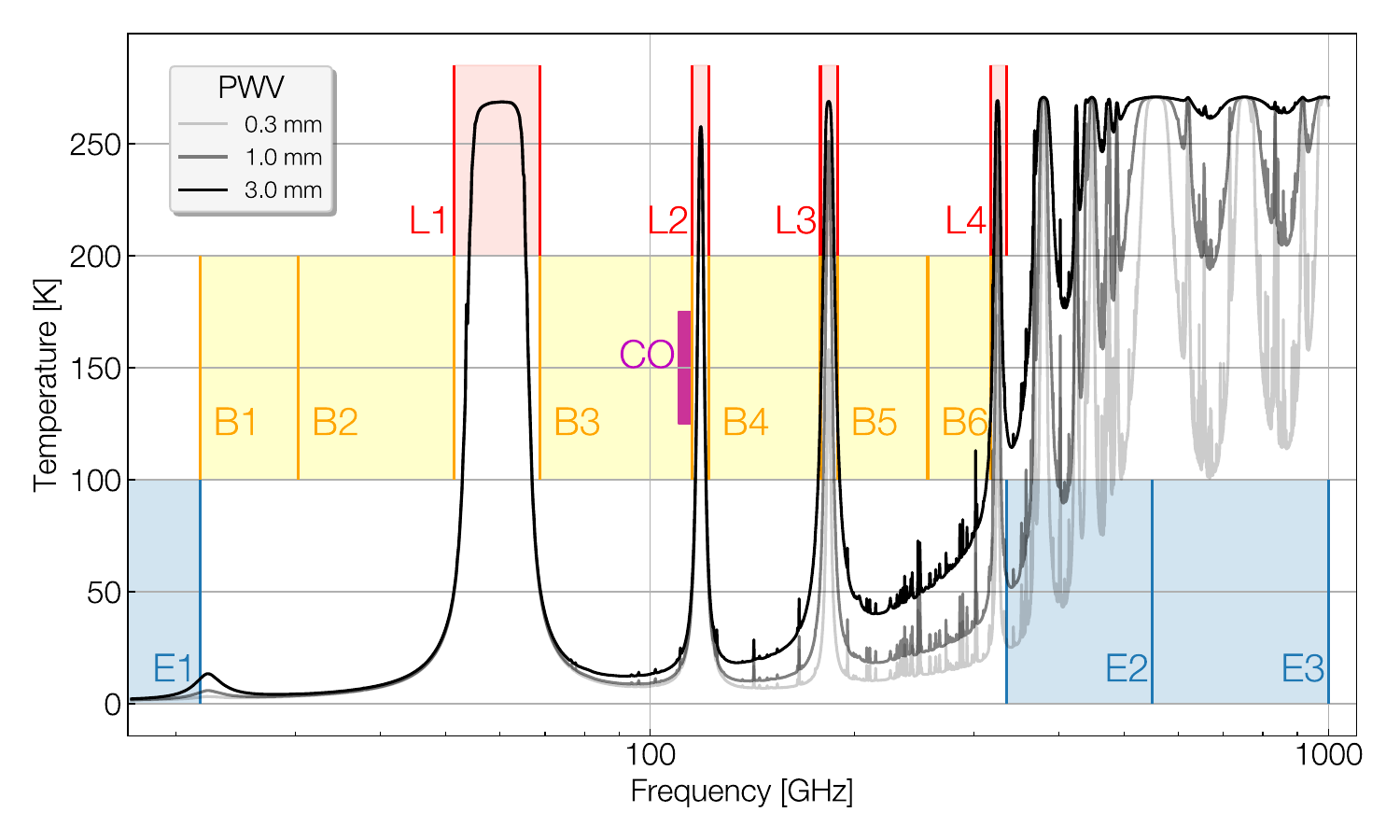}
    \caption{The atmospheric emission expected at the Chilean site, modeled using the \texttt{am} code \citep{paine_2023_8161272}. This emission is shown for increasing values of precipitable water vapor (PWV). The highlighted regions show the spectral regions we use to evaluate out-of-band leakage in Table \ref{tab:blueleakresults}: atmospheric lines (row 1), in-band regions (row 2), spectral end regions (row 3), and the galactic J = 1 $\rightarrow$ 0 CO transition line.}
    \label{fig:atmosphere}
\end{figure*}

The passbands of our detectors are chosen to efficiently detect light at frequencies within the relevant atmospheric transmission windows and to substantially suppress light at frequencies where the atmospheric emissivity is nearly saturated, thereby minimizing noise on the detectors. Figure \ref{fig:atmosphere} shows the spectrum of atmospheric emission expected at the site in Chile, modeled using the \texttt{am} code~\citep{paine_2023_8161272}. Regions of high emissivity are highlighted as crucial spectral regions which must be avoided by the end-to-end passband of the instrument.

In planning the measurement program, we defined a set of parametric requirements across the full range of frequencies that could be observed by the instrument. This leads to a relatively large number of parameters to verify, but has the advantage of relying completely on passband parameters rather than a separate calculation of atmospheric impact for each measurement. The key parameters to constrain for a complete validation of the passband are band-shape, center frequencies, location of the band edges, and control of out-of-band leakage.

The first regions to consider are the target bands (labeled B1-B6 in Figure \ref{fig:atmosphere}) which are the frequency ranges in-band for the various detectors that make up Simons Observatory. Using B3 as an example, this region would correspond to the 90 GHz passband. The in-band requirements specify that the center frequency and bandwidth of B3 must not deviate from its target design by more than 5\%. The coupling in all other regions that are considered out-of-band to B3 (B1, B2, B4, B5, and B6) are defined by the threshold constraint that the resulting decline in mapping speed is no more than 2.5 percent for the neighboring bands (in this case, B2 and B4) and 0.5 percent for the other bands more widely separated from the main band. The targets for the other detector channels are defined in the same manner.

Next, we consider regions occupied by emission lines from oxygen and water vapor in the atmospheric foreground (L1-L4), and carbon monoxide in the galactic foreground (CO). We use an instrument sensitivity model (Section \ref{sec:on-sky}) to calculate the impact that signal leakage from these regions would have on the final mapping speed. We set benchmarks for the acceptable amount of leakage from these line regions to be a fractional amount of the total band power, with this fractional amount constrained to bound the reduction in mapping speed to less than 2.5\% for atmosphere lines bordering an in-band region (e.g., L1 and B3) and less than 0.5\% for all other line-band pairs (e.g., L1 and B4). As anticipated, the coupling in most of these spectral regions was well below the benchmark (Table \ref{tab:blueleakresults}). 

The final bands we consider are the ``spectral end regions'' that lie beyond the Simons Observatory science bands (labeled E1, E2, and E3). For E2 and E3, the requirement is set to limit the decline in mapping speed from atmospheric contamination to less than 0.5 percent. For the lower spectral end region E1 where the atmospheric emission is small, the priority is to constrain radio-frequency interference (RFI) pickup on the detectors.
\begin{figure*}
    \includegraphics[width=\textwidth]{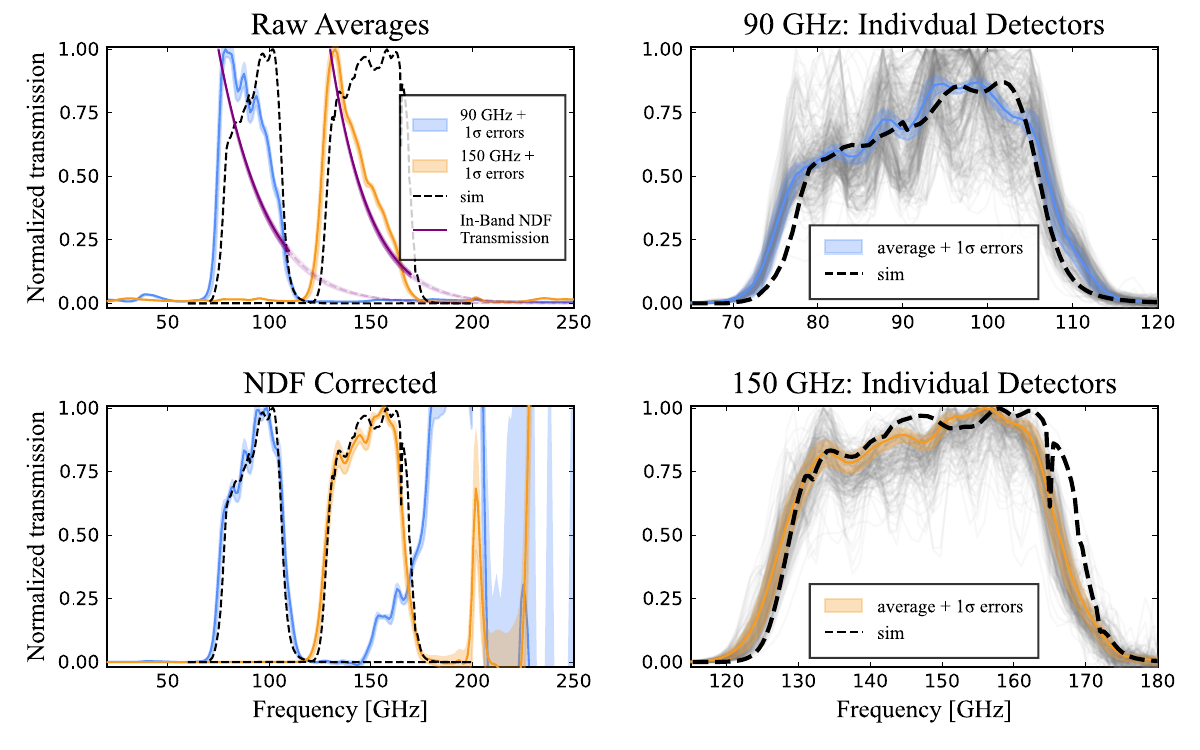} \caption{\textit{Top left}: Raw averages of all 90 GHz (blue) and 150 GHz (orange) detectors with $1 \sigma$ statistical error bars. Simulated bands are shown in black.
    \textit{Bottom left}: Final passband averages of all detectors measured with the FTS. The effects of the NDF are corrected for by dividing out its calibration function. This correction has the side-effect of exponentially magnifying the noise floor, which introduces a large bias at high frequencies. To remove this bias, we subtract an exponential fit term based on estimates of the noise floor. The error bars are a quadrature combination of $1 \sigma$ statistical and NDF calibration function errors. The main band shapes are in excellent agreement with the passbands expected from simulations of the designed optical system, while any spurious above band features and noise are exponentially magnified by the NDF correction. 
    \textit{Right}: All 265 90 GHz and 195 150 GHz FTS detector measurements are plotted together. The simulated bands and final averages are over-plotted. These measurements have been corrected for the NDF transmission.}
    \label{fig:band_averages}
\end{figure*}

\subsection{\label{sec:band_methods} Measurements}
Constraining the spectral response across the whole range of bands described above requires measurement of detector behavior from close to 0 GHz to 1 THz. Covering this frequency range necessitates the use of multiple complimentary methods.

An FTS is used for in-band measurements and for measurements at lower frequencies. A set of focusing optics at the output of the FTS allows its illumination to be localized to about a dozen pixels on the focal plane at any single XY-position above the window, with an average signal-to-noise ratio over 100. The systematics of the FTS and coupling optic system are well understood to the 2 percent level across the in-band region~\citep{Alford_2024}. However, the FTS measurement can couple detector non-linearities into spurious harmonics of the primary passband. Additionally, the fundamental limitations of the FTS signal-to-noise ratio constrains its ability to detect very small signals. This is especially limiting at frequencies higher than the in-band region due to the exponential attenuation of the signal by the NDF. Measurements of the single-pixel box, with its high saturation point, allow us to extend this range.  

Complimenting the FTS measurements is a broad-band (20-1200 GHz) coherent frequency-selectable narrow-band source generated from the interference of two infrared lasers, which we refer to as the FLS. It is coupled via free space frequency-independent attenuators which allow the signal to be varied from tens of $\mu$W to pW enabling in band measurements and high signal to noise out-of-band measurements. These are complementary to the FTS as they have different systematic effects where there is overlap and can extend beyond the capabilities of the FTS for small signals.

The out-of-band leakage is additionally constrained with measurements using thick grill filters~\citep{Timusk:81} and large metal meshes. Thick grill filters consist of metal plates with precision drilled holes that block radiation at low frequencies but efficiently pass radiation above a cutoff that is defined by the waveguide cutoff of the holes. These are used to verify the out-of-band leakage at high frequencies and check the band-edge regions. At the lowest frequencies we use metal meshes (similar to fencing) to block radio frequencies and set limits on the sensitivity to radio-frequency interference.

The following paragraphs explain these measurement methods and their analyses procedures in greater detail.

\subsubsection{Fourier-transform Spectrometer (FTS)}
The FTS used in these measurements is a Primordial Inflation Explorer-style FTS~\citep{PIXIE} originally created for calibration of the Atacama Cosmology Telescope passbands\citep{Swetz_2011}. The FTS output is sent through coupling optics which matches the FTS output to the beam of the optics tube. The distance between the FTS output and the secondary focus of the optics tube is about 30~cm. The coupling optics focal ratio is  matched to the large aperture telescope receiver optics tubes with ratio f/2.4~\citep{Gudmundsson:21}. The coupling optics focus the FTS output to a small subset (25-50) of detectors. Thus the FTS was mounted in-lab to motorized XY-stages which allowed us to scan the coupling optics output over the entire optics tube window. Detector array measurements consisted of a 5x5 grid of passbands measured over the focal plane to ensure that all detectors were well illuminated by the field of view of the FTS. Only one FTS position was needed to measure the single pixel detectors.

The FTS data are taken by fixing the FTS XY-position, moving the FTS scanning mirror to a specific position, taking data for several seconds with the input blackbody source chopped at 8 Hz, demodulating the signal at this frequency by fitting the time series for a sine and cosine component, adding the resulting amplitude in quadrature to get an estimate for the signal, then stepping the phase delay mirror and repeating data acquisition to build up an interferogram at that particular spatial position. This approach leads to a small additive bias from squaring the noise which must be removed in the analysis.

At any given XY-position, the FTS illuminates multiple pixels on the focal plane. As a result, each detector is typically measured several times throughout the full measurement scan. For each detector, the measurement with the highest signal-to-noise ratio is selected for use in the final band averages, since these correspond to the measurement with the best FTS coupling to that detector. Next we average together the bands with a weight given as the square of the signal-to-noise ratio. The NDF transmission is corrected for by dividing out the fitted transmission function from the final averages. Finally, since this NDF correction causes the noise bias to become frequency dependent, we subtract an exponential term to de-bias the final interferogram. The fit for this exponential term ignores the in-band data.

\subsubsection{Frequency-selectable Laser Source (FLS)}
The FLS~\citep{fls2024} is a coherent source capable of emitting a wide frequency range with a narrow line width. The source uses two infrared diode lasers and a transmitter photomixer to generate signals between roughly $15$~GHz and $1.2$~THz~\citep{TopticaManual}. The source outputs about $100~\mu$~W at 100~GHz, which is far beyond the saturation limits of the detectors. Therefore, coupling this source to the detector arrays requires significant attenuation in addition to what is added by the NDF. A lesser degree of attenuation is required to operate the high-$G$ single pixel box.

Attenuation is achieved by reflecting the emitted beam, which has been collimated by an off-axis parabolic mirror, off of up to five removable attenuating prisms, each of which attenuate the signal power by a frequency-independent factor of $1/20 $~\citep{fls2024}. The prisms are made of high-density polyethylene. Their geometry is designed to reduce scattering from inside the prism to re-enter the main beam. Reflectivity measurements of the prisms show a neutral response across frequency. Swapping mirrors with the prisms allows us to adjust the power incident on the detectors by a factor of up to a million, providing a high dynamic range for the measurement of faint signals (e.g., in the out-of-band regions). After the light travels through the attenuating assembly, it is focused into the optics tube with an off-axis parabolic mirror followed by a lens that matches the f-number of the optics tube. The FLS and attenuating assembly are mounted on the motorized XY-stages on our instrument mounting structure. This allows the FLS to move over the cryostat window at different positions.

The FLS, which is single-moded, illuminates several pixels at a time, though we choose to focus on the ones that have a decent beam after completing a beam map. A measurement consists of running with a high-attenuation prism configuration to probe the in-band response, and subsequent measurements at lower attenuation to measure the out-of-band regions of the passband. The resulting data sets are re-scaled by the relevant attenuation factors to produce a spectrum which is NDF corrected (if needed) and normalized to match the in-band response of the FTS. Since the source is beam-filling and single moded, no frequency dependant corrections are needed to compare with the FTS data. 
 
\subsubsection{High-pass filters}
A high pass filter can be used to probe leaks at frequencies higher than the in-band region. At high frequencies, the simplest high pass filter is the thick grill filter, which uses a grid pattern of holes machined on a thick metal plate to block frequencies below the designed cutoff frequency~\citep{Timusk:81}. For these measurements, an array of thick grill filters with different cutoff frequencies were designed and machined to fit on a plate that is placed over the LATR-Tester window and prevents any unfiltered light from entering the optics tube. To probe the detector response, a thermal source is positioned directly over the filter opening and chopped to generate a modulated signal with a known frequency cutoff. To correct for filter transmission loss and confirm the cutoff frequencies, each filter was measured in-lab using the FLS.

Additionally, high pass filters with much lower frequency cutoffs are used to constrain any effects of RFI pickup in the detector data. To achieve this, a wire mesh cage was installed in front of the cryostat window with aluminum tape sealing around the edge of the wire mesh. The wire mesh had 0.25" square openings, which creates a high pass filter at around 5 GHz, blocking radiation from sources such as Wi-Fi and Bluetooth communications that are prevalent in region E1.

\subsection{\label{sec:in-band} Passband Measurements}
Measurements of the passbands were taken in Cooldowns 1, 2, and 4 using a Fourier Transform Spectrometer. Cooldowns 1 and 2 were used to calibrate the NDF transmission by using the single pixel measurements with and without the NDF, while Cooldown 4 repeated measurements to obtain data for the final optics tube configuration (without the 1~K low-pass edge filter). The results of this section are reported from Cooldown 4, which reflect the final optical configuration of the optics tube.

The band averages before and after NDF correction are shown on the left side of Figure~\ref{fig:band_averages}. The estimated error ranges include both statistical errors and errors from the fitted NDF transmission function, summed in quadrature. Final measurement counts contain 77\% (295) of working 90 GHz detectors and 95\% (195) of working 150 GHz detectors. Individual passband measurements, corrected for the NDF, are shown on the right side of Figure~\ref{fig:band_averages}. The main band shapes and calculated central frequencies are in excellent agreement with predictions, with measured central frequencies of $92.9 \pm 1.0$ GHz and $148.0 \pm$ 1.1 GHz and simulated centers of 92.7 and 149.5 GHz.

\begin{figure}[ht]
    \includegraphics[width=.47\textwidth]{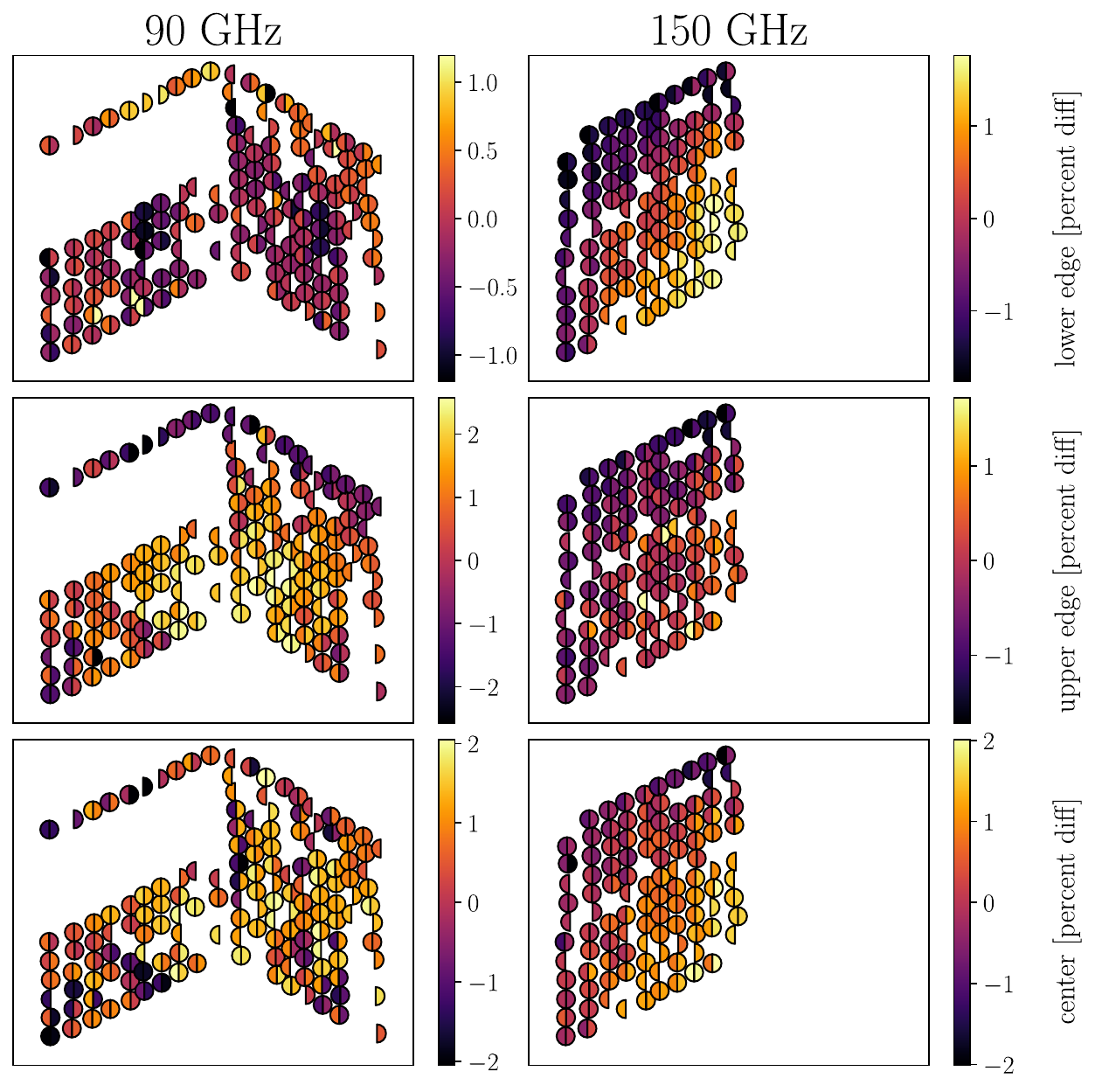}
    \caption{Lower edge, upper edge, and center frequencies of individual detector passbands plotted vs focal plane position. The focal plane positions are found by mapping through the coupling optics of the FTS to obtain finely focused beams. The lower and upper edges are fitted as 5\% transmission frequencies. To reduce noise in the fitted edges, these passbands are not corrected for the NDF transmission. There is a clear radial trend in the band edge frequencies, with pixels at the wafer edge measuring about 3\% lower frequency than pixels at the wafer center. This is more prominent in the upper edge of 90 GHz, lower edge of 150 GHz, and upper edge of 150 GHz, all of which are defined by on-chip filters. The lower edge cutoff of the 90 GHz detectors is defined by the waveguide response that is independent of wafer properties. `A' group detectors are plotted with a left half-circle, and `B' or missing group detectors are plotted with a right half-circle. The large sections of missing detectors are due to a loose-fitting coaxial connection on the prototype detector array which has since been reworked for the production arrays. The colorbar percent spread shown is bounded by 99\% intervals for the dataset to avoid outlier effects.}
    \label{fig:band_attrs}
\end{figure}

Passband attributes for all detectors on the array are examined in Figure~\ref{fig:band_attrs}. The large sections of missing detectors on this wafer map are noted to be caused by a loose-fitting coaxial connection on the prototype detector array, which has since been reworked for the production arrays. The upper and lower band edges are calculated by fitting the passband data to find frequencies with 5\% normalized transmission. 

There are several interesting observations in these passband statistics. First, the edges of the band appear to have a radial dependence, with a total gradient of about +3\% from edge pixels to center pixels. This radial variation is an expected but not entirely understood systematic effect of the dielectric wafer. Next, it is clear that the low edge of the 90 GHz passbands do not exhibit this radial trend. This is also consistent with design expectations, where that lowest edge is fixed by the waveguide cutoff rather than the on-wafer circuitry. Finally, there are no significant differences in center frequency or bandwidth between paired detectors in the same pixel.

\subsection{\label{sec:leakage}Out-of-Band Constraints}
Constraining out-of-band leakage across all regions of interest spanning 1 to 1000~GHz requires a combination of FTS, FLS, and thick grill filter measurement. The leakage requirements for each region, outlined in Table\ref{tab:blueleakresults}, are defined as a percentage threshold of out-of-band power relative to total in-band power for a frequency-independent source.

\paragraph{FTS} 
Measurements taken with the FTS are used to constrain all out-of-band and atmospheric line regions at frequencies below the 90 / 150 GHz passbands to within the baseline specification, as shown in Table \ref{tab:blueleakresults}. At higher frequencies, these constraints are strongly hindered by the NDF correction which grows exponentially with frequency and leads to massive uncertainties in the detector array measurements for frequencies beyond the 90 / 150 GHz in-band regions (as seen in Figure \ref{fig:band_averages}). Therefore, any FTS constraints at these higher frequencies are measured with the high-saturation single-pixel box which does not require an NDF correction. Due to its low optical efficiency and high levels of thermal pickup, these higher frequency single-pixel box constraints should be taken as strict upper bounds. All other regions are constrained using the detector array average passbands described in Section~\ref{sec:bandpass}. The out-of-band leakage values are obtained by integrating the measured passbands over the target out-of-band region and normalizing by the integral over all frequencies, with a cutoff at high frequencies to avoid spurious effects from the NDF correction. 

\begin{figure}
    \centering
    \includegraphics[width=1\linewidth]{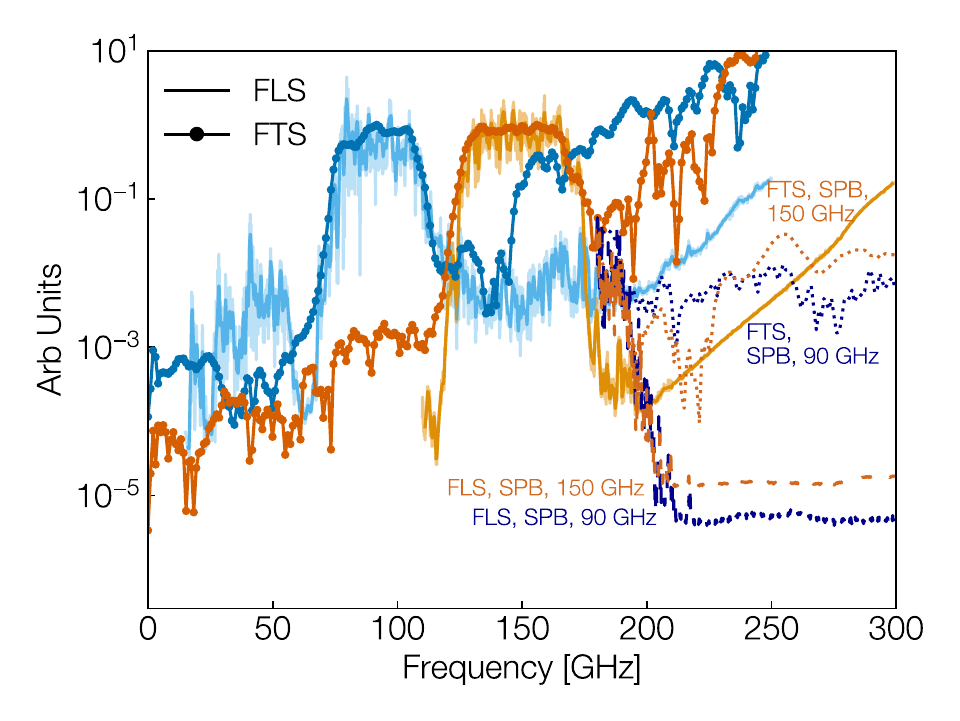}
    \caption{ Passbands for individual channels on the detector array (solid / scatter) and the single-pixel box (dashed / dotted) measured using the FLS and the FTS. The variable power of the FLS allows for a larger dynamic range in the out-of-band regions than is possible with the constrained signal-to-noise ratio of the FTS. Additionally, comparing the FLS and FTS passbands together helps to highlight out-of-band features in the FTS spectral response that are due to instrumental systematics. The single-pixel box, or SPB, is plotted for high frequencies where the detector array measurements are limited due to the NDF correction. These SPB measurements highlight the powerful ability of the FLS to probe significantly smaller signal levels than the FTS.}
    \label{fig:fls_ufm_bands.png}
\end{figure}

\paragraph{FLS}
The spectral regions measured using the FTS were repeated using the FLS and extended to higher frequencies. Due to the very small illumination of the FLS, passband measurements must be performed on a per-detector pixel basis. This limits the total sampling of passbands to just a few detector channels across the array, in contrast to the FTS method which can more efficiently sample the full detector array. Figure~\ref{fig:fls_ufm_bands.png} shows a comparison of FLS and FTS passbands measured on the same detector channels. As was the case with the FTS measurements, the high-frequency regions of the detector array bands are dominated by uncertainty in the NDF correction. However, the greater dynamic range of the FLS over the FTS is clearly observed in these $>200$ GHz out-of-band regions. This cross-check is also essential in identifying instrumental systematics that are specific to the FTS setup. Additionally, the FLS was used for measurements using the single-pixel box, with the higher frequency response also being shown in Figure~\ref{fig:fls_ufm_bands.png}. Since no NDF correction is required for the single-pixel box, these measurements show the full dynamic range improvement that is possible with the FLS over the the standard FTS method. At the high out-of-band regions, the FLS can probe signals nearly 3 orders of magnitude below what is achievable with the FTS.

Table~\ref{tab:blueleakresults} provides a summary of fractional leakage in the atmospheric and out-of-band regions resulting from the FLS. The leakage percentage is determined by integrating the passband within the region of interest and normalizing it against the integral of the passband across the entire measurement range. In most cases, we use individual detector passbands from the detector array for these calculations. However, due to the exponential increase in NDF correction with frequency, we use measurements from the single-pixel box detectors for some of the higher frequency regions. It is important to note that the single-pixel box detectors exhibit low optical efficiency and high thermal pickup. Therefore, the passbands and subsequent out-of-band leakage calculations derived from these detectors represent upper bounds. Since the FLS measurements were only performed up to 300~GHz, the value presented for Region B6 corresponds to a smaller frequency range spanning 256.6~GHz to 300~GHz, rather than the nominal 256.6~GHz to 317.4~GHz. The mismatch of leakage values between the FLS and FTS measurements are in part due to the larger sample size of detectors available with the FTS method.

The FLS passbands are Hanning smoothed over a width of 1~GHz. These smoothed passbands are used for the leakage values calculated in Table~\ref{tab:blueleakresults}, although the unsmoothed passbands yield similar estimates. The baseline specifications for each spectral region are met using the FLS measurements, with the exception of Region B4 which shows higher-than-expected coupling from the 90~GHz channels into the 150~GHz in-band region. Since this is not an atmospheric line region, there is no concern over mapping speed degradation. There is an additional requirement that pertains to the J = 1 $\rightarrow$ 0 CO line, which is at the high edge of the 90~GHz band and covers 110-115~GHz. The CO line leakage requirements are defined as a peak-normalized transmission amplitude. The FLS passband measurements yield respective transmission amplitude values of 12.5\% and 1\% at the low and high edges of this requirement.

\paragraph{High-pass filters}
Thick grill filters are used to place upper bounds on the fractional leakage for every out-of-band region at frequencies higher than the 90 / 150 GHz passbands (Table~\ref{tab:blueleakresults}). For the highest frequency regions $>$ 300~GHz (L4, B6, E2, and E3), these upper bounds are effective at constraining leakage to within the baseline specifications. For L4 and B5, which are closer to the detector passbands, a tighter constraint is necessary. To achieve this we use two filters with cutoffs at each edge of the target region. For example, the L4 atmospheric line region is constrained by designing two filters with cutoffs at 178 and 189 GHz each. The difference in detector response between these two filters corresponds to the expected power in this region.

To probe the E1 region, we look for the presence of glitches and other artifacts in the detector responses with and without the Faraday cage installed over the window. The data without the Faraday cage demonstrates a significant number of timestream glitches. These glitches primarily occur every 8 seconds with a smaller group of large signal glitches occurring every 120 seconds. Both categories of glitches were reduced by more than 90\% when the Faraday cage with a cutoff at about 5 GHz was installed over the window. This test indicates that the detectors in the LATR-Tester cryostat are sensitive to RFI pickup, in this case likely coming from the building wireless network. Tests at the Chilean site will be necessary to determine whether this will remain an issue after deployment.
 
\begin{deluxetable*}{ccccccccccc}
    \label{tab:blueleakresults}
    \caption{Out-of-band leakage measurements and design specifications for each spectral region illustrated in Figure~\ref{fig:atmosphere}.}
    \tablehead{
        \colhead{} & \multicolumn{2}{c}{Edges (GHz)} & \multicolumn{2}{c}{Baseline spec (\%)\tablenotemark{a} } & \multicolumn{2}{c}{FTS (\%)} & \multicolumn{2}{c}{TGF (\%)} & \multicolumn{2}{c}{FLS (\%)}\\
        \colhead{Spectral Region} & \colhead{Low} & \colhead{High} & \colhead{90~GHz} & \colhead{150~GHz} & \colhead{90~GHz} & \colhead{150~GHz} & \colhead{90~GHz} & \colhead{150~GHz} & \colhead{90~GHz} & \colhead{150~GHz}
        }
    \startdata
        \multicolumn{2}{l}{Atmospheric lines\tablenotemark{b}} \\ \hline
        Line L1&51.4 &68.8 &1.0 &0.4 & 0.08&0.01&-&-&0.15&-\\
        Line L2&115.4&122.0& 0.8 &1.0 & 0.2&0.2&-&-&0.51&0.04\\
        Line L3&178.2&188.8&0.2 &0.9 &-&$<$0.6&0.7&0.7&0.27&0.02\\
        Line L4&317.4&335.0&0.1 &0.4 &-&-&$<$2e-3&$<$3e-3&-&-\\ \hline
        \multicolumn{2}{l}{Out-of-Band Response} \\ \hline
        Region B1&21.7&30.3&1.0 &1.0& 0.1&0.7&-&-&0.04&-\\
        Region B2&30.3&51.4&1.0 &1.0 &0.6&0.94&-&-&0.42&-\\
        Region B3&68.8&115.4&N/A &1.0 &N/A&2.6&N/A&-&N/A&-\\
        Region B4&122.0&178.2&1.0 &N/A &-&N/A&-&N/A&1.82&N/A\\
        Region B5&188.8&256.6&1.0 &1.0 &1.9&2.1&0.5&0.3&0.17\tablenotemark{c}&0.09\tablenotemark{c}\\
        Region B6&256.6&317.4&0.6 &0.9 &1.6&4.3&$<$1e-3&$<$1e-3&$<$1e-3\tablenotemark{c,d} &$<$0.01\tablenotemark{c,d}\\ \hline
        \multicolumn{2}{l}{Spectral End Regions\tablenotemark{e}} \\ \hline
        Region E2&335.0&550.0&0.05 &0.07 &-&-&$<$1e-3&$<$3e-3&-&-\\
        Region E3&550.0&1000.0&0.03 &0.03 &-&-&$<$1e-3&$<$3e-3&-&-\\
     \enddata
     \tablenotetext{a}{Leakage values are defined as a fractional percentage of out-of-band power relative to the total in-band power for a frequency-independent source.}
     \tablenotetext{b}{The J = 1 $\rightarrow$ 0 CO line avoidance region spans 110-115~GHz. The CO line leakage requirements are set as the peak-normalized transmission amplitude rather than as a fraction of integrated power. Using the FTS, the lower (110~GHz) and upper (115~GHz) edge transmission values are 26\% and 2\%. Using the FLS, these transmission values are 12.5\% and 1\%. The baseline specifications are set at 10\% and 2\%, respectively. The impact of partially missing this design target is not yet fully understood, but we do not expect a major systematic effect.}
     \tablenotetext{c}{Value from single pixel box measurement, single detector.}
     \tablenotetext{d}{Leakage calculated in the region defined from 256.6 GHz to 300 GHz as the FLS measurement went up to 300 GHz.}
     \tablenotetext{e}{Region E1 is not constrained by a fractional power threshold for this work.}
\end{deluxetable*}

\section{\label{sec:beams} Beam Characterization}
The beam emerging from the optics tube must satisfy a number of requirements. It must correctly couple to the telescope to produce the correct beam size on the sky. It must be well-controlled by the optics such that stray radiation which spills past the telescope to the ambient temperature telescope structure does not compromise the instrument's sensitivity. Finally, the detailed beam pattern should be checked for evidence of optical components with unexpected behavior. Here we present a program of detailed laboratory measurements to check these properties. These measurements use both the traditional thermal source method~\citep{Cothard_2018} as well as near field holography~\citep{chesmore2022}.

Radio holography takes advantage of the antenna theory relationship: the far-field radiation pattern of a reflector antenna is the Fourier Transformation of the field distribution in the aperture plane of the antenna~\citep{alma_holog}. Using Fresnel diffraction~\citep{Goodman2005-ne}, the measured fields can be propagated through the optical system to determine the stray power spilling outside of the mirrors of the telescope and the far-field beam pattern of the telescope fed by this receiver. These measurements enable a detailed verification of system-level optical performance prior to the deployment of a receiver on a telescope. This represents a significant advance in the ability to characterize optical systems prior to deployment. The thermal measurements are presented as confirmation of the holography results and to explore the impact of the NDF on optical performance in the test configuration.

\subsection{\label{sect:holog-method} Holography Beam Map}
Tunable monochromatic millimeter-wave sources are used to measure the beam properties at 5~GHz intervals over the band from 70–170~GHz. These signals are broadcast into the receiver using standard gain feedhorns held 10~cm from the the cryostat window on a motorized two-dimensional stage. The signal is detected using a special purpose receiver in the focal plane comprising a harmonic mixer coupled to a feedhorn identical to those used in the detector arrays. This measurement is carried out with the focal plane at 4~K and does not require the NDF. The results fully characterize the optical performance of the system.

During a measurement, the source frequency is fixed, and the source is moved over a $50$x$50$~cm range in 0.25~cm steps. Complete measurement details, including open-source hardware and software for optical modeling, are described in~\cite{chesmore2022}. This results in a measurement of the amplitude and phase of the beam.

\begin{figure*}[t]
    \centering
    \includegraphics[width=.85\textwidth]{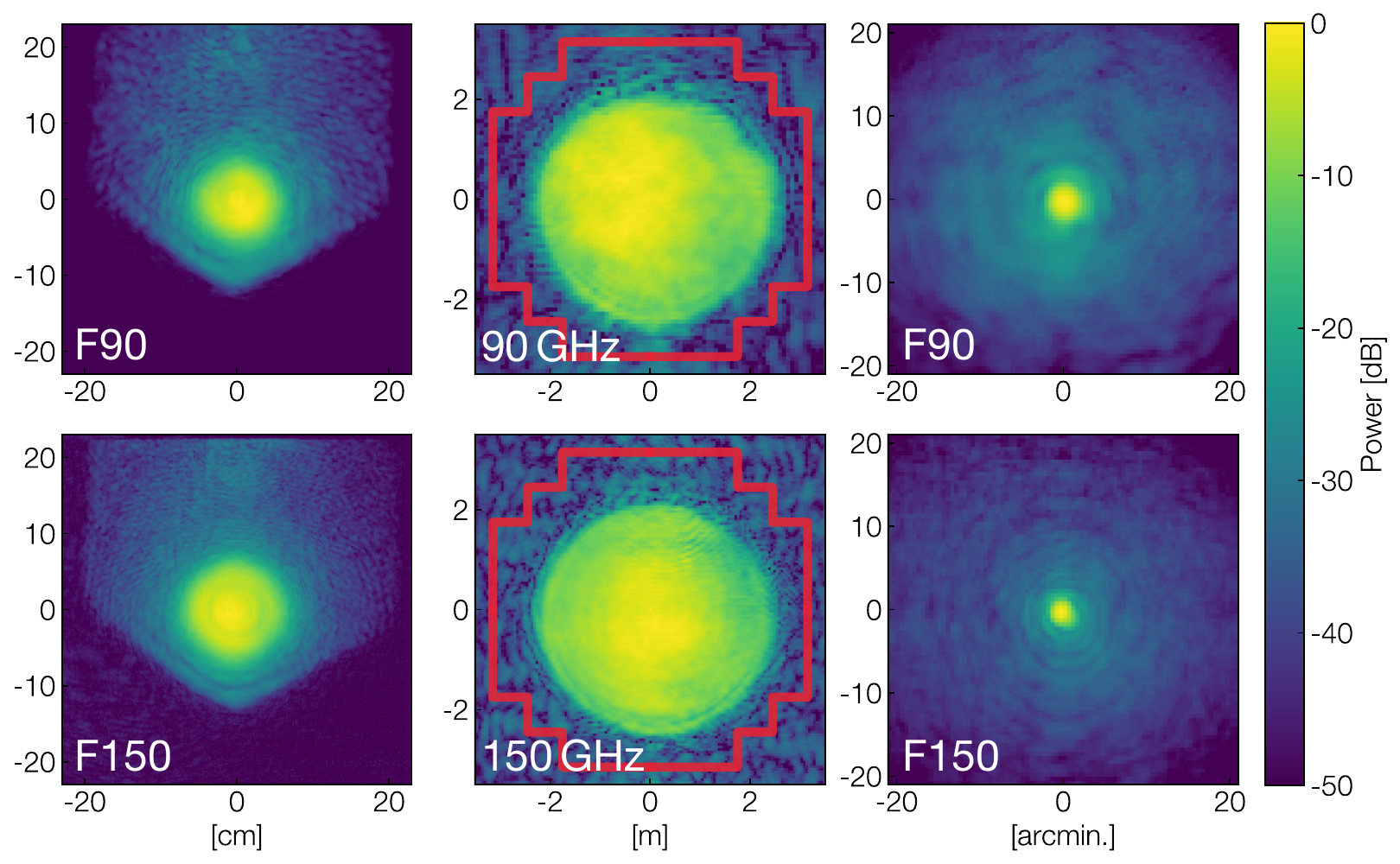}
    \caption{\emph{Left column}: Holography beam map measurements in the mid-frequency band.  \emph{Middle column}: All beam maps propagated at the secondary illumination of the large aperture telescope.  The spilled power to 300~K is calculated by integrating power outside the boundary of the secondary (red line) with respect to total integrated power of the map.  \emph{Right column}: Using Fresnel diffraction, the LATR-Tester measurements are propagated through the large aperture telescope from the near-focal plane to the far-field.}
    \label{fig:holo}
\end{figure*}

The average beam response at frequencies within the 90~GHz and the 150~GHz bands are shown in the left column of Figure~\ref{fig:holo}. These represent the fields emerging from the window of the cryostat. The bright, roughly Gaussian structure is the expected main beam, while the hexagonal signal which contains roughly 2.5\% of the power of the beam represents scattering and other non-ideal effects from filters, lenses, and other components. The impact of this scattering must be assessed through its impact on scattering beyond the telescope and on the main beam shape.  

\paragraph{\label{sect:secondary-illumination}  Fields at Plane of the Secondary Mirror}
To determine the amount of power ``spilled'' to 300~K, we propagate the measured fields onto the plane of the secondary mirror, approximately 12~m away from the measurement plane. This is accomplished by using the Fourier relationship between the near-field $E(x,y)$ and far-field $B(\theta_x,\theta_y)$ beams~\citep{McIntosh2016,alma_holog},

\begin{equation}
    B(\theta_x,\theta_y) = \int_{\text{aperture}} E(x,y) e^{ i \frac{2\pi}{\lambda} (\theta_x x + \theta_y y )} dx dy \ ,
\end{equation}

\noindent where the complex electric field $E(x,y)$ is integrated over the area of the aperture, and $\lambda$ is the wavelength.

The middle column of Figure~\ref{fig:holo} shows beam maps propagated to the secondary mirror, with the boundary of the secondary mirror in red. To quantify the spilled power, the total power outside of the boundary is integrated and then normalized to the total integrated power of the beam map. This yields an average spilled power of 0.65\% and 0.68\% in the 90 and 150~GHz bands, respectively, with no significant frequency dependence. This is below the design target of 1\% and indicates that the sensitivity of the instrument should not be compromised by spilled power to 300~K. For reference, if the warm spillover had instead been 3\% as was measured on the similar ACT receiver~\citep{Cothard_2018}, it would have reduced mapping speed of the instrument by 30\%. As this was a key performance driver, this is an important constraint on the instrument performance. 

\paragraph{Far-Fields}\label{sec:far-fields}
To propagate the measured near-fields into the far-field, a virtual telescope is used to produce the fields from a distant (100~km) point source on the measurement plane. The resulting far-fields are shown in the right hand column of Figure~\ref{fig:holo}. A detailed comparison of the far-field holography measurement to simulations is presented in~\cite{chesmore2022}. These comparisons show that the holography data are consistent at the few percent level with the predicted 90~GHz~(150~GHz) full-width half maximum of 2.16~(1.38)~arcmin and with low ellipticity and no unexpected features such as the ``little-buddies'' seen in the Atacama Cosmology Telescope~\citep{2021arXiv211212226L,Gudmundsson:21}.

\paragraph{Filter Removal}
\label{sec:filter}

\begin{figure}[ht]
    \centering
    \includegraphics[width = .45\textwidth]{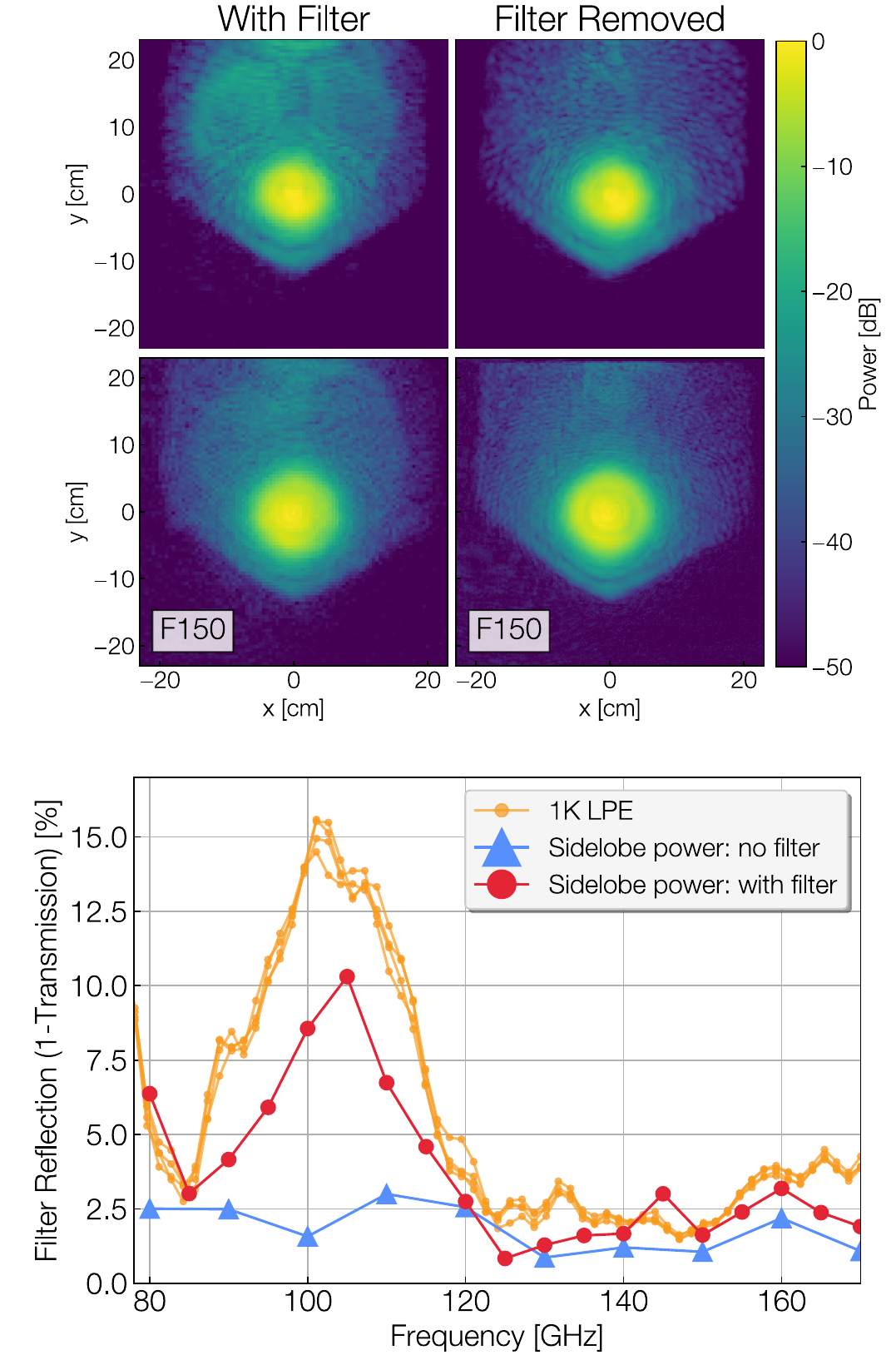}
    \caption{\textit{Top}: Band-averaged near-field beam maps with and without the 1~K low-pass edge filter (LPE) in the optics tube. With the filter, beam maps show extra scattering in the top parts of the map, through the upper portion of the LATR-Tester hexagonal window (hexagonal power around the main beam at roughly -20\,dB due to reflection). \,\,\textit{Bottom}: In orange is the measured reflectivity (1-transmission) of the 6.8\,cm$^{-1}$ low-pass edge filter, which is placed at the 1~K stage in the optics tube.  The red (blue) line shows the measured integrated fractional power outside the main beam at each frequency with(without) the filter in the optics tube.}
    \label{fig:filter_info}
\end{figure}

The holography results described so far are from the final LATR-Tester optics configuration (Cooldown 4). In an earlier optics tube configuration (Cooldown 3) which included an additional filter (1K low-pass edge filter in Figure~\ref{fig:latrt}), holography measurements revealed extra scattering outside of the main beam at the window (hexagon at $\sim-20$~dB) which was as high as 10\% in the 90~GHz band (Figure \ref{fig:filter_info}). While these side-lobes did not significantly change the spilled power to 300~K, they would have reduced optical efficiency and led to an enhancement of near side-lobes of the on-sky beam.

The frequency dependence of side-lobe power is quantified by integrating the fractional power outside of the main beam in the raw holography data measured at the window. The radius of the main beam is defined as 13.5~cm and corresponds to a $-20$~dB drop in power. Figure~\ref{fig:filter_info}B shows the integrated fractional power outside the main beam as a function of frequency, and also the measured reflectivity of the 1~K 6.8~cm$^{-1}$ low-pass edge filter~\citep{Cardiff_2006}.

By comparing the side-lobe power response to the reflectivity measurements of each filter in the optics tube, a close match is found in the response of the 1~K low-pass edge filter. A Zemax~\citep{zeemax} simulation is used to investigate the effect that a reflective filter at the 1~K stage would have on the optics tube beam. This simulation predicts that rays which reflect off the filter will end up outside of the main beam in the form of extra side-lobes, similar to what was observed in the holography measurements.

It is notable that the amplitude of the integrated power closely matches the amplitude of reflections from the filter. As the integrated signal requires a bounce from the filter and a bounce from the horn array at the focal plane, this implies that the horn arrays are $\sim 75\%$ reflective. 

After removing the low-pass edge filter from the optics tube, a holography measurement shows a clear decrease in side-lobe power. Figure~\ref{fig:filter_info}A and B shows the reduced near-field side-lobe power following the removal of the filter, and the comparison of the side-lobe power with the filter in place. With the filter removed, the new side-lobe fractional power across the band is 1.9\%, a factor of 3 lower than with the filter installed. The tests presented above do not include this filter in the optical chain. As a result of these measurements, the 1~K low-pass edge filter has since been removed from the nominal design of the LATR optics tubes. This provides a concrete example of how holography measurements can be used to optimize cryogenic optical systems.

\subsection{\label{sec:thermal_beams}Thermal Beam Maps}
A second option for measuring beams out of the optics tube is to scan a chopped thermal source above the LATR-Tester window to directly modulate the response of the Simons Observatory detectors at the focal plane. Unlike the holography measurements, thermal beams are an incoherent measurement – there is no phase information and the overall signal-to-noise ratio is lower. The important distinction is that these thermal beam maps measure the response of the detectors rather than the holography receiver. Therefore, consistency between thermal beams and holography beams serves as an important validation of the results.

\paragraph{Thermal source}
The thermal source is a ceramic IR lamp that reaches a maximum stable temperature of 480~\degree C. The lamp sits inside a light pipe with a 45~\degree angled mirror that allows the lamp to operate in a horizontal configuration, which was necessary for proper thermal regulation of the system. The modulation of the thermal source is achieved using a 14" diameter spring steel chopper blade with four cutout notches positioned in front of the thermal source. The fan blade is coupled to a servo motor that is capable of rotating the blade up to a constant 60~Hz velocity, or an effective 240~Hz modulated signal. An iris at the output of the chopper enclosure allows the source size to be adjusted between 5 and 75~mm.

\paragraph{Measurement procedure}
To map the beam pattern of the detectors, the source is chopped at 8~Hz while scanning an XY-grid at 24~cm above the window. The choice of 8~Hz as the modulation frequency avoids contamination from spurious harmonics, such as those from pulse tube cooler vibrations and is well within the optical time constant of our detectors. The high signal-to-noise of the detectors means that full beam maps of every channel with 0.5~cm resolution can be made in only a few hours of measurement time. In a 300~K lab environment and behind the carefully-optimized NDF, the detectors achieve a signal-to-noise level of about $\sim 40$~dB. 

\begin{figure}
    \centering
    \includegraphics[width=0.95\linewidth]{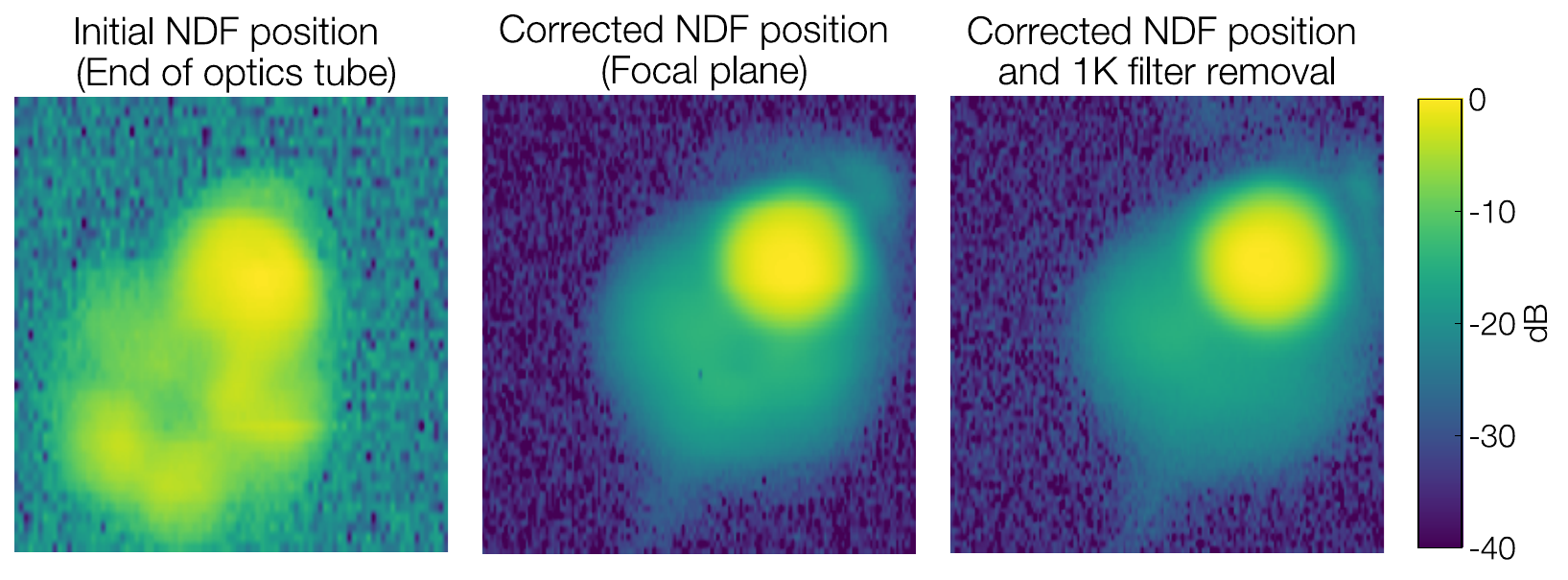}
    \caption{A single detector beam across three separate cooldown configurations. The first beam map (CD2) shows significant reflections and spill off the copper NDF mounting plate after it was initially installed at the end of the 4~K optics tube. The second beam map (CD3) shows a much improved beam quality and signal to noise after repositioning the NDF to 20~mm from the focal plane. In the third beam map (CD4), the 1~K low pass filter is removed with no adverse impact to the beam shape or noise level. The diffuse background hexagon observed in these beam maps is due to reflections between the copper focal plane and the NDF, and will not be present in the on-sky optical configuration.}
    \label{fig:beam_evo}
\end{figure}

Besides demonstrating the design predictions, these thermal beam maps are also used to reveal unexpected structures in the optics and confirm features shown with holography characterization. As shown in Figure \ref{fig:beam_evo}, thermal maps taken in Cooldown 2 with the NDF installed at the end of the optics tube showed a surprising ghosting pattern with a hexagonal shape that bears a striking resemblance to the NDF mount. This pattern is believed to be an effect of reflections from the copper NDF mount and caused a significant degradation in the quality of the detector beams. These thermal maps informed the decision to relocate the NDF to a position just above the focal plane, thereby removing the unwanted pattern in subsequent cooldowns. Thermal beam maps are also used to confirm that the removal of the 1~K low pass filter, which previously caused excess sidelobe power in the holography beam maps, does not negatively impact the behavior of the detector beams.

\section{\label{sec:polarization}Polarization Performance}
Cross-polarization, where linearly polarized detectors are optically sensitive to the polarization orthogonal to their designed polarization, leads to a reduction in polarization efficiency of the system. This reduces the effective optical efficiency for polarization measurements and so a complete characterization of the optical efficiency requires constraining the level of cross-polarization in the optical system. 
Similar to the beam characterizations, the polarization performance was tested using two methods: a measurement of cross-polarization in the optics using holography, and a measurement of cross-polarization in the detectors using a polarized thermal source.

\subsection{Holography Cross-Polarization}
\begin{figure}
    \centering
    \includegraphics[width = .45\textwidth]{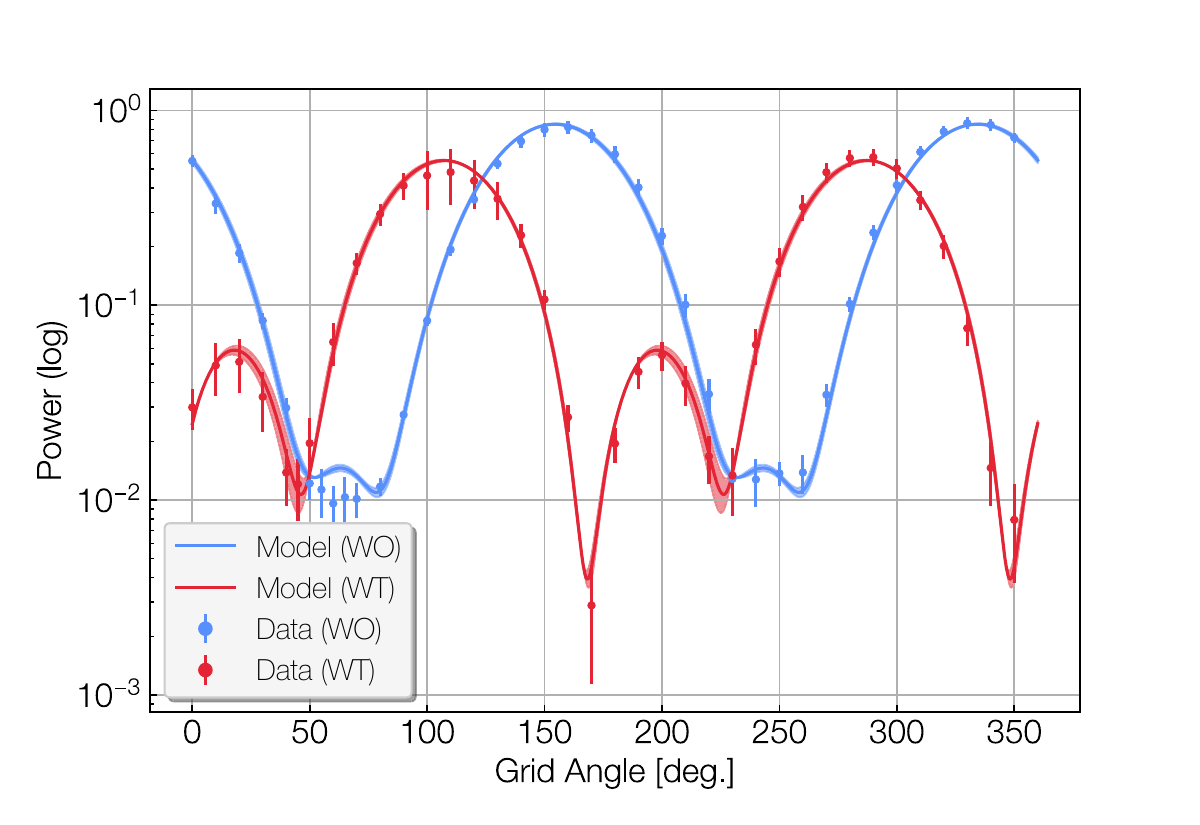}
    \includegraphics[width = .45\textwidth]{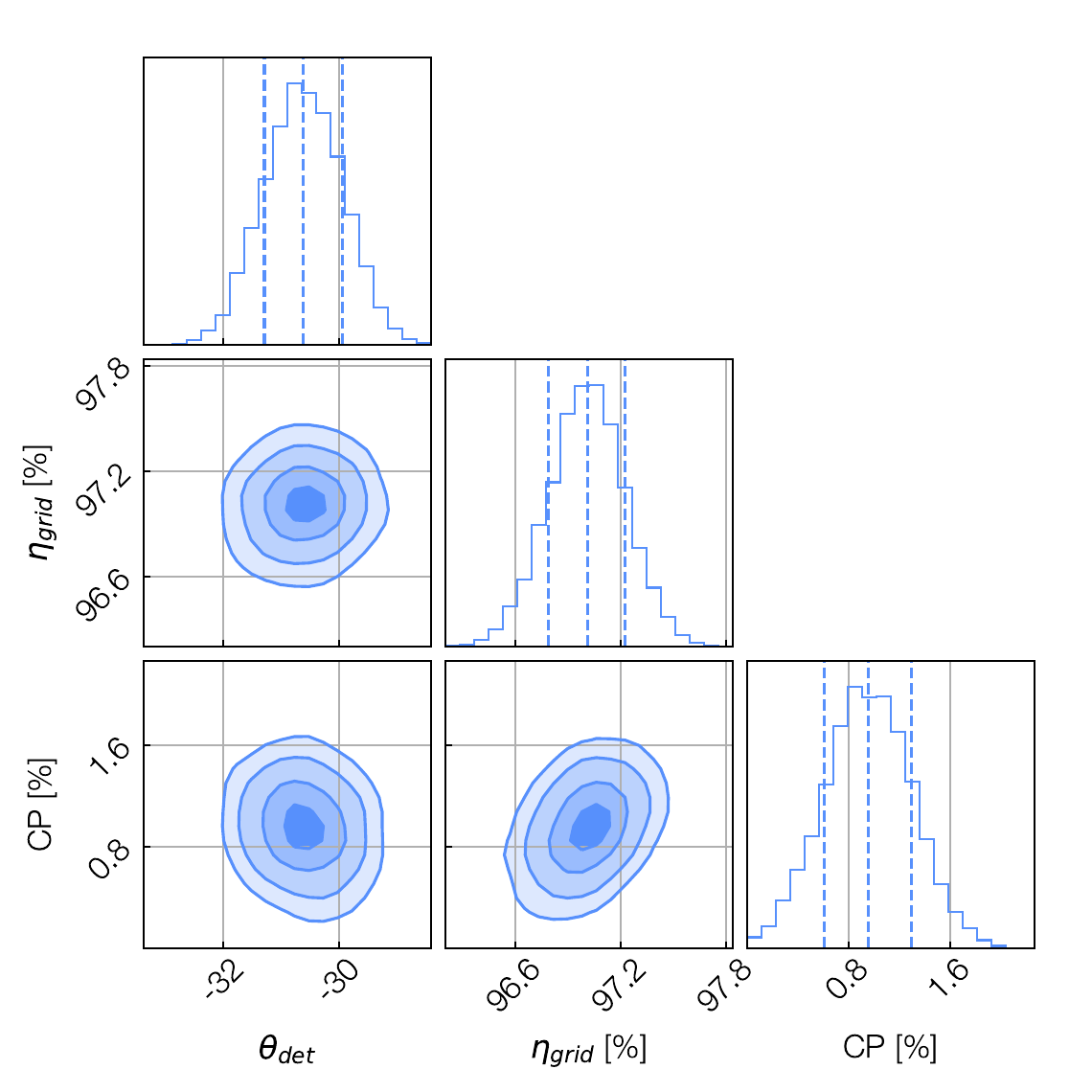}
    \caption{\textit{Top}: Polarized beam power (band-averaged from 85-120~GHz) as a function of grid rotation, measured with the holography setup described in Section~\ref{sect:holog-method}. Co-polarization\,(WO, blue) holds the source waveguide aligned with the receiver orientation. Cross-polarization\,(WT, red) uses a $90^{\circ}$ waveguide to make the source aligned perpendicular to the receiver orientation. The polarization model \citep{Harrington_Polarization_Simulation_Software_2022} is fit using an MCMC with no instrument polarization. \,\, \textit{Bottom}: Constrained parameters from the polarization model include cross-polarization $\mathrm{CP}$, receiver angle $\theta_{\text{det}}$, and grid efficiency $\eta_{\text{grid}}$. The cross-polarization of the optical system is constrained to the percent level.}
    \label{fig:holo_crosspol_params}
\end{figure}

The holography system described in Section~\ref{sect:holog-method} uses a linearly polarized source and a linearly polarized detector. It is straightforward to insert a waveguide twist into the setup to rotate the source polarization by $90^\circ$. A cross-polarization measurement can be carried out by inserting a polarizer into the optical path, rotating it over a range of angles, rotating the source polarization and repeating the measurement.  

During installation, the polarization angles of the source and receiver were intentionally misaligned at an angle of about $30^\circ$ to simplify the coaxial paths inside the optics tube. Outside the cryostat, the source can be operated with or without a $90^{\circ}$ twist waveguide, enabling the injection of two orthogonal linear polarizations. The measurements without the twist waveguide are denoted as ``WO'' and measurements with the twist waveguide are denoted as ``WT.'' Neither are aligned directly along the angle of the receiver inside the cryostat.

A thin-film linearly polarizing ``grid'' that could be rotated through $360^\circ$ was mounted in between the holography source and the window of the cryostat. Separate measurements made on the detector array with this thin-film polarizer and a separate wire grid polarizer constrain the polarization efficiency of the thin-film polarizer to $97 - 99\%$ for the 90~GHz band. The response of the holography receiver was measured as a function of the angle of the polarizing grid with and without the twist waveguide installed on the source at 5~GHz increments between 85 and 120~GHz. The band averaged response for these measurements is shown in the top plot of Figure~\ref{fig:holo_crosspol_params} where both the WO and WT responses are normalized to the maximum of the co-polar measurements. 

The data from these holography measurements are fit to a comprehensive polarization model that incorporates a Mueller matrix formalism to predict the Stokes parameters of the system~\citep{Harrington_Polarization_Simulation_Software_2022}. We marginalize over nuisance parameters to account for uncertainties in the measurement, such as the angle of the waveguide twist. The result is robust constraints on the holography receiver angle, the wire grid efficiency, and the cross-polarization. Figure~\ref{fig:holo_crosspol_params} shows the constrained parameters obtained from the fits to this polarization model.  The receiver angle $\theta_{\text{det}}$ is constrained to $-30.62^{\circ\,+0.69}_{-0.67}$, the grid efficiency $\eta_{\text{grid}}$ is constrained to $97.00\%^{+0.21}_{-0.22}$, and the cross polarization $CP$ is constrained to $0.96\%^{+0.34}_{-0.34}$. 

\subsection{Detector Cross-Polarization}
The holography polarization setup constrains the cross-polarization of the combined optical components inside the optics tube, but it does not impose any constraints on detector cross-polarization. A measurement of the polarization performance of the detector array was attempted using two setups. In the first, a rotating wire grid polarizer is mounted flat in front of a chopped thermal source and positioned directly above the beam position on the window. In the second, a thin-film polarizer is mounted at an angle on the window. In both cases, the polarized grid is illuminated by unfocused light over a broad range of angles of incidence. These tests implied an unexpectedly-high cross-polarization at the 10\% level across all channels on the array, with dependencies that include wafer position, frequency, and polarization angle of the detectors.

An obvious issue with this measurement is that the radiation pattern emerging from the source is not well-controlled or well-matched to the beam of the optics tube. To mitigate this potential systematic, a new measurement of detector polarization has been made on an ultra-high-frequency (220 / 280~GHz) optics tube using a focused and well-controlled beam. The results of this optically-focused test, to be detailed in a forthcoming publication, demonstrate percent-level cross-polarization performance across the detector array. In contrast, a repeat of the unfocused measurement on the same 220 / 280~GHz optics tube yielded a similar high limit of approximately 10\%. For this reason, it is believed that the intrinsic cross-polarization of the optics tube is best estimated with the holography measurements. It is noted that on-sky measurements using a similar camera on the Atacama Cosmology Telescope show similar percent-level cross-polarization performance. Future cross-polarization measurements should use focused beams tailored to match the receiver's input with well-controlled polarization properties.

\section{Integrated Detector Performance}
\label{sec:det_perform}

\begin{figure}[t]
    \centering
    \includegraphics[width=\linewidth]{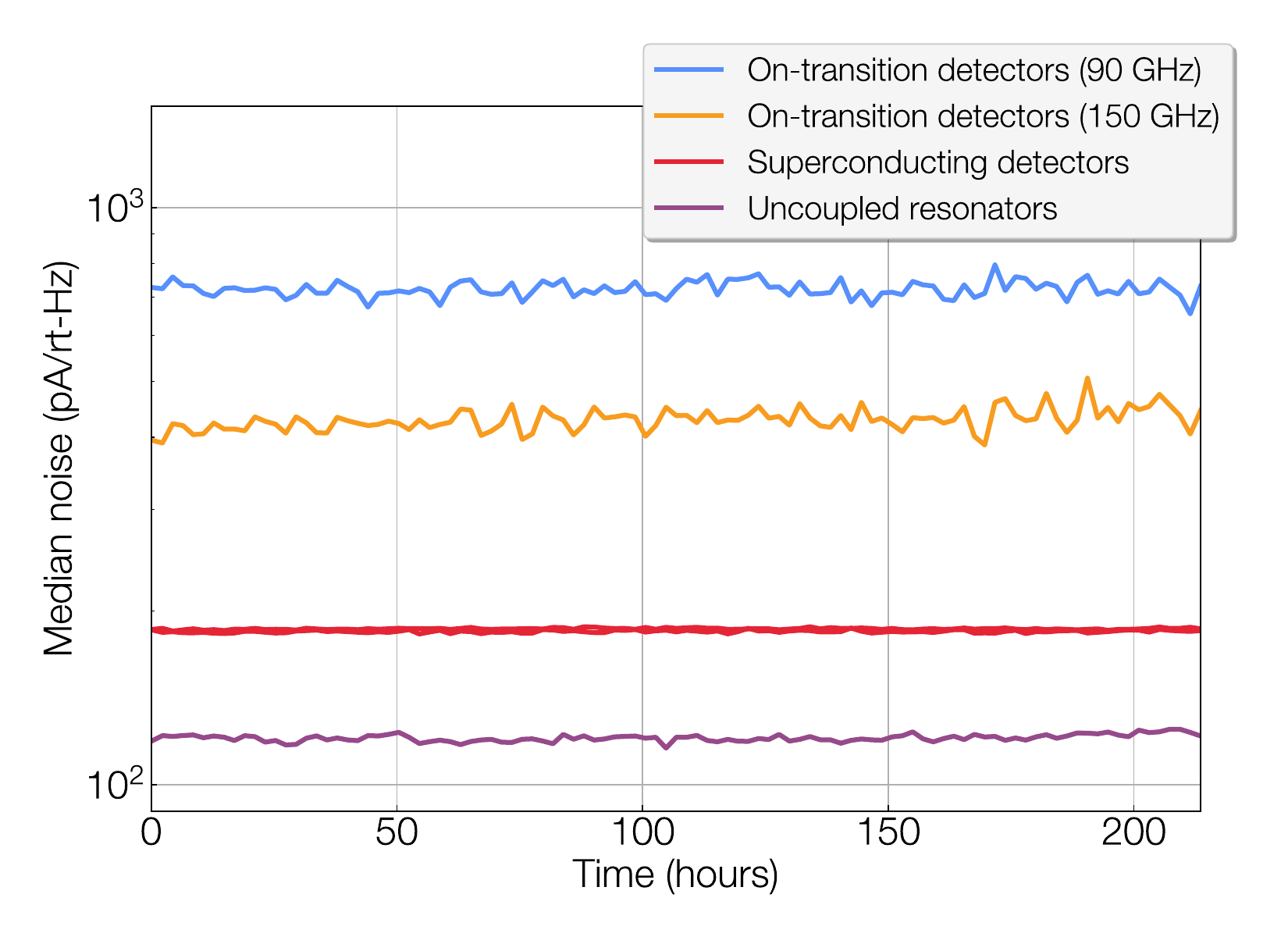}
    \caption{\label{fig:det_performance} {White noise curves taken over the span of 9 days in a noisy lab environment demonstrate the long-term stability of the readout system. Plotted channels include uncoupled readout resonators along with coupled superconducting and on-transition detectors. Note that the absolute noise levels plotted here are not indicative of the expected field performance.}}
\end{figure}

Detector performance metrics such as noise levels, time constants, and readout stability are also measured on the LATR-Tester. On their own, these measurements represent an accurate picture of the expected performance of a detector array integrated with the full optics and readout electronics. When compared with previous component-level testing, these measurements serve as powerful cross-checks across the separate Simons Observatory testing facilities. The testing pipeline for the Simons Observatory focal-plane modules begins with cold screening of the microwave multiplexing components to validate performance quality~\citep{Whipps_2023,Huber_2022}. Fully-integrated detector modules are then tested inside of a dark cryogenic testbed~\citep{McCarrick2021_MFUFM,yuhan}. Once tested, these focal-plane modules are integrated into the LATR-Tester for a final end-to-end optical validation. Consistency across all of these testing stages is a robust test of the expected detector array performance.

Detector parameters as measured with the LATR-Tester, including saturation power, critical temperature, and thermal conductance, are consistent with the testing results from the dark tests performed for the same detector array, up to known calibration discrepancies. These tests are performed in a similar ``dark'' LATR-Tester configuration with a metal plate covering the window and a beam-filling cold load at the end of the 4~K optics tube (see Section \ref{sec:cold_load}). Heaters on the focal plane allow the temperature of the 100~mK stage to be adjusted, which is used to probe the saturation powers of the detectors across different thermal bath temperatures. The agreement of these detector parameters across institutions is an important indicator that the detectors and readout system are behaving in a consistent and expected manner. In addition to these ``dark'' tests, the LATR-Tester has the capability of testing two other detector effects which require optical loads external to the cryostat: long-term noise stability in an ambient environment and optical time constants.

\subsection{Noise Stability}
The high sensitivity of the detector arrays makes them highly susceptible to instabilities when coupled to the outside environment. RF interference, temperature drifts, electrical power fluctuations, and other environmental effects can cause spikes or longer-timescale instabilities in the readout noise.

Long-term readout stability of the deployed instrument enables seamless data acquisition and a high observing efficiency, and is therefore a crucial metric for gauging the performance of the system. On the LATR-Tester, the long-term noise stability of the readout was tested by recording $\sim$5 minutes of detector data every hour for over 10 days in a 300~K lab environment. Figure \ref{fig:det_performance} shows the median detector noise level of these observations averaged from 5 to 10 Hz with an exceptionally stable response across the full duration of the test. This test includes channels with bare resonators (i.e., no coupled detectors) as well as channels with both superconducting and on-transition detectors.

\subsection{Optical Time Constants}
The time constant of a CMB detector is a crucial parameter that must balance telescope scanning speed with stability of the detector readout system. Typically, time constants are characterized in-lab by observing the thermal decay of biased detectors in response to small voltage square waves imposed on the detector bias lines~\citep{Koopman_2018}. This bias step response time can act as a proxy for the optical time constant, which is then measured on-sky after instrument deployment.

\begin{figure}
    \centering
    \includegraphics[width=\linewidth]{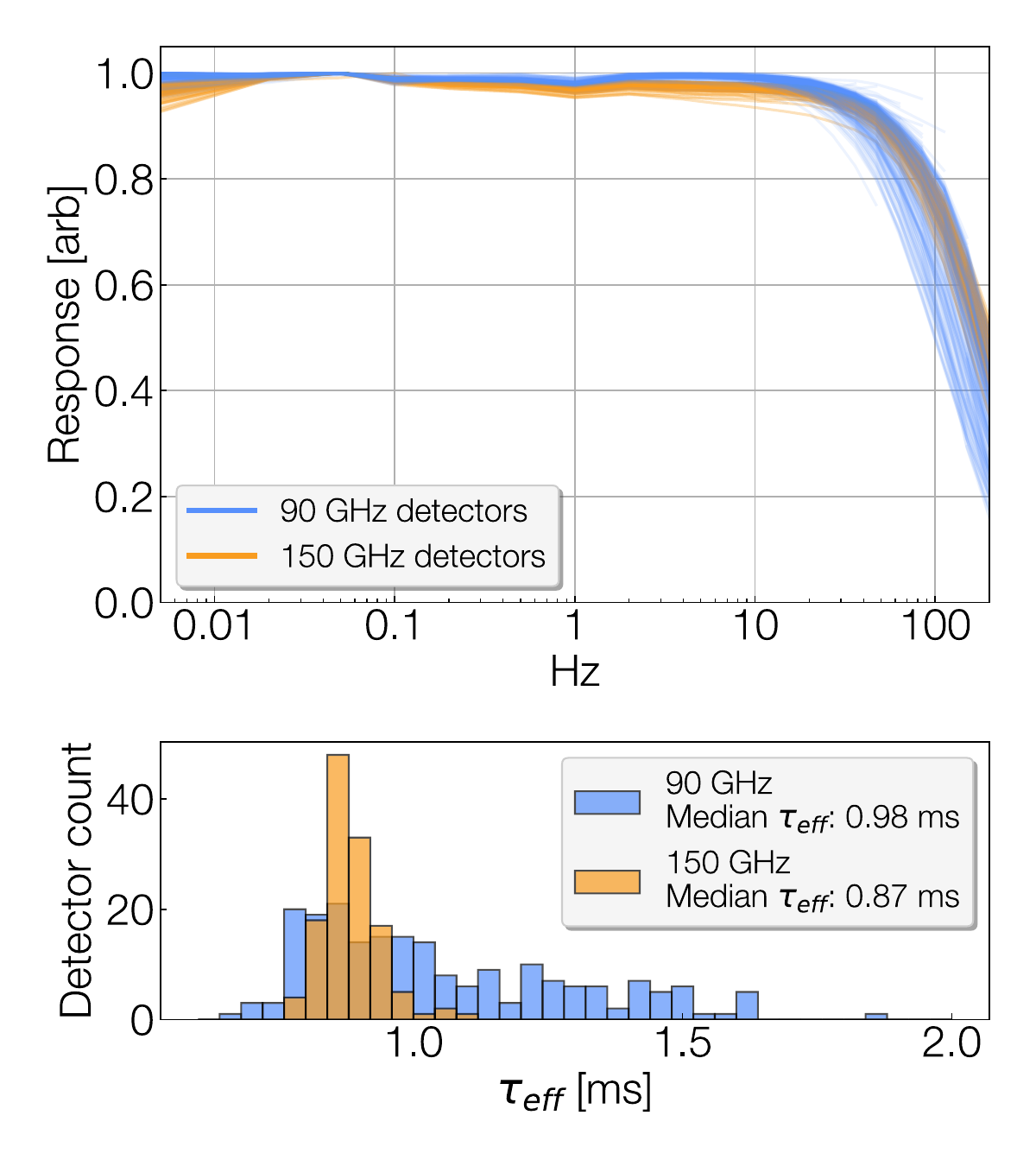}
    \caption{\textit{Top}: Optical response of detectors to a thermal source chopped between 5~mHz and 200~Hz. The response is plotted in arbitrary units and fit to a single pole low pass filter model to obtain an effective time constant, $\tau_\mathrm{eff}$ for each detector channel. \textit{Bottom}: Spread of the effective time constant fit parameter measured across the full detector array.}
    \label{fig:taus}
\end{figure}

With the LATR-Tester, direct in-lab measurements of the optical time constants are possible using the same chopped thermal source used for thermal beam mapping (Section \ref{sec:thermal_beams}). The stability and large frequency range of the chopped source allows the modulation response of each detector to be tested from as low as 5~mHz up to 200~Hz. For this measurement, the source is positioned above the detector beam and slowly stepped up in chopper frequency to obtain a demodulated response at each frequency step. The effective time constants are obtained by fitting this data to a single pole filter response function (Figure~\ref{fig:taus}). It should be noted that the effective time constant is dependent on the level of optical loading on the detectors. At 90~GHz, the optical loading through the LATR-Tester matches the average loading expected from the atmosphere in Chile, after taking into account attenuation through the NDF.

The chopped source frequency response of each detector channel and their associated time constant fits, $\tau_\mathrm{eff}$, are shown in Figure~\ref{fig:taus}. The median effective time constants are measured to be just under 1~ms for both the 90 and 150~GHz channels. These time constant measurements meet the target specifications set by the scan rate of the large-aperture telescope at the 90 and 150 GHz beam widths. The larger spread in $\tau_\mathrm{eff}$ that appears in the 90~GHz detectors can be partially attributed to variances in optical loading, which are compounded by the frequency dependence of the NDF. Finally, we note that time constants taken using the bias steps method under the same loading conditions are somewhat consistent with optical time constants, trending about 10–20\% slower than the optically-chopped measurement but with similar spreads.

\subsection{Readout crosstalk}
An important performance metric of any detector readout system is the level of spurious correlated signals observed between any two channels. This so-called crosstalk can be a major systematic effect to the on-sky analysis effort if not properly controlled. The impact of this effect is expected to grow as the packing density of detector pixels increases, making crosstalk a significant concern for next-generation CMB experiments. In the microwave multiplexing scheme employed by SO, the primary source of crosstalk arises from a pairwise parasitic coupling of the resonators in frequency-adjacent channels, though other mechanisms are known to exist at lower amplitudes~\citep{groh2023crosstalk,mates_2019}.

The LATR-Tester presents a unique opportunity to constrain readout crosstalk in an optical environment. Using the FTS coupling optics, we can probe the response of a single detector channel independently from its frequency-adjacent neighbor. Any observed response in the nearest-neighbor channel would correspond to a crosstalk signal. To avoid potential optical systematics, we consider only frequency-adjacent channels that are not also physical neighbors on the focal plane. In addition to measuring a crosstalk signal, we can perform a beam map with this coupling optics setup and search for correlated ``ghost'' beams for each detector channel. The appearance of a ghost beam at the position of the detector channel's nearest-neighbor is then a spatial characterization of the observed crosstalk for that detector pair.

The detector pairs in this analysis are selected by finding all frequency-adjacent channels with physical beam centers that are separated from each other by more than 5~cm. This scanning distance corresponds to the approximate width of a single beam and is chosen to avoid physical overlap between any two beams. This filtering has the effect of cutting about half of the total nearest-neighbor candidates on the array. The crosstalk between neighboring detector pairs is determined by multiplying the pixel values of the two beams and integrating the resulting correlation beam within a small 3.5~cm radius circle to avoid overlap with the main beam sidelobes. The resulting sum is then normalized by squaring the integration of the original beam, providing an estimate of the correlated amplitude of the two test beams relative to the amplitude of each individual beam.

\begin{figure}
    \centering
    \includegraphics[width=1.0\linewidth]{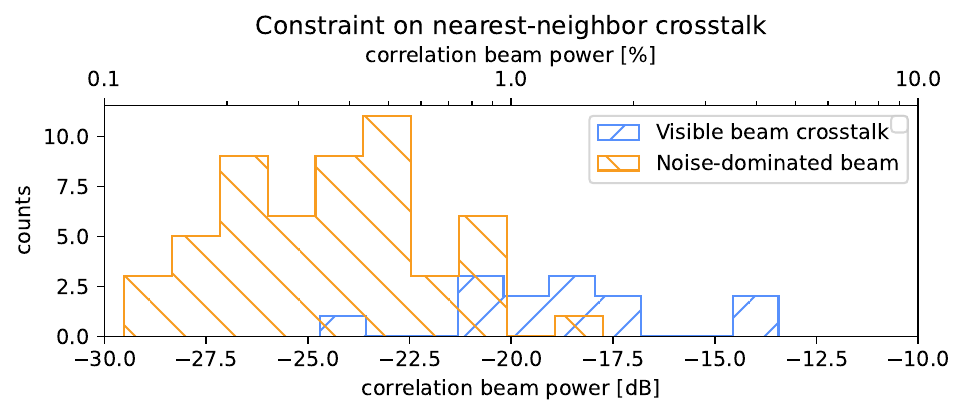}
    \vspace{-0.2cm} 
    \includegraphics[width=1.0\linewidth]{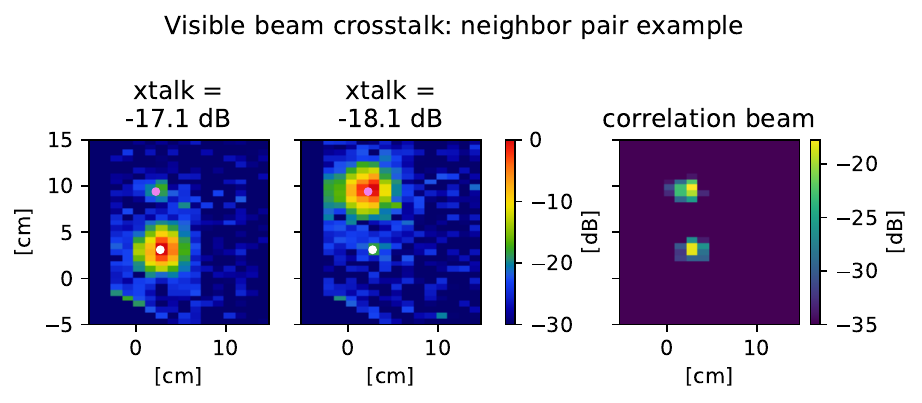}
    \caption{\textit{Top}: Readout crosstalk on the detector array is constrained by measuring the signal power within correlation beams of frequency-adjacent resonator pairs. This measurement is limited by a scattering noise floor at the $20-25$~dB level; therefore, the majority of these correlation beam measurements should be interpreted as an upper bound constraint on readout crosstalk to below 1\%. Channels with visible ghost beams indicate a true crosstalk signal that is measurable above this noise floor. \, \,
    \textit{Bottom}: Example of a neighboring pair of detector channels with relatively high electrical readout crosstalk of about 1\%. A ghost beam is observed in each beam map directly near the center of the beam of the nearest-neighbor resonator, with dots indicating the center of the detector beams for reference. The amplitudes of the correlation beams are used to estimate crosstalk for these channels.}
    \label{fig:crosstalk}
\end{figure}

Figure \ref{fig:crosstalk} shows an example of these measured ghost beams for a single pair of nearest-neighbor channels, and the distribution of signal power within the correlation beam for a total of 33 channel pairs, or 66 total channels. We note that this measurement is strongly limited by noise and scattering off of the detector hexes at the $20-25$~dB level, which constrains our ability measure crosstalk signals with an amplitude below the noise floor. Therefore, the correlation beam power we measure for a majority of these 66 channels is consistent with noise and should be interpreted as an upper bound constraint ($<1\%$) on the actual readout crosstalk. On the other hand, we interpret channels with ghost beams that are visible above the noise floor as indicative of true readout crosstalk. The amplitude of these crosstalk signals appear to be largely driven by resonator spacing, with lower crosstalk observed for channels that are spaced further apart in frequency.

Finally, we note that the effects of crosstalk nonlinearity~\citep{groh2023crosstalk} are not well-constrained in this analysis. There are two competing $\mathcal{O}(1)$ effects due to this nonlinearity which are unmodeled by our analysis; thus these measurements can be considered a proxy but not a direct measurement of the crosstalk expected during observations. Future iterations of this measurement will be enhanced with longer integration times to improve signal-to-noise and map resolution, which will enable a more complete analysis of readout crosstalk and its nonlinear effects.

\section{\label{sec:detector-gain}Detector Responsivity Calibration}
\begin{figure*}[t]
\centering
    \includegraphics[width=\textwidth]{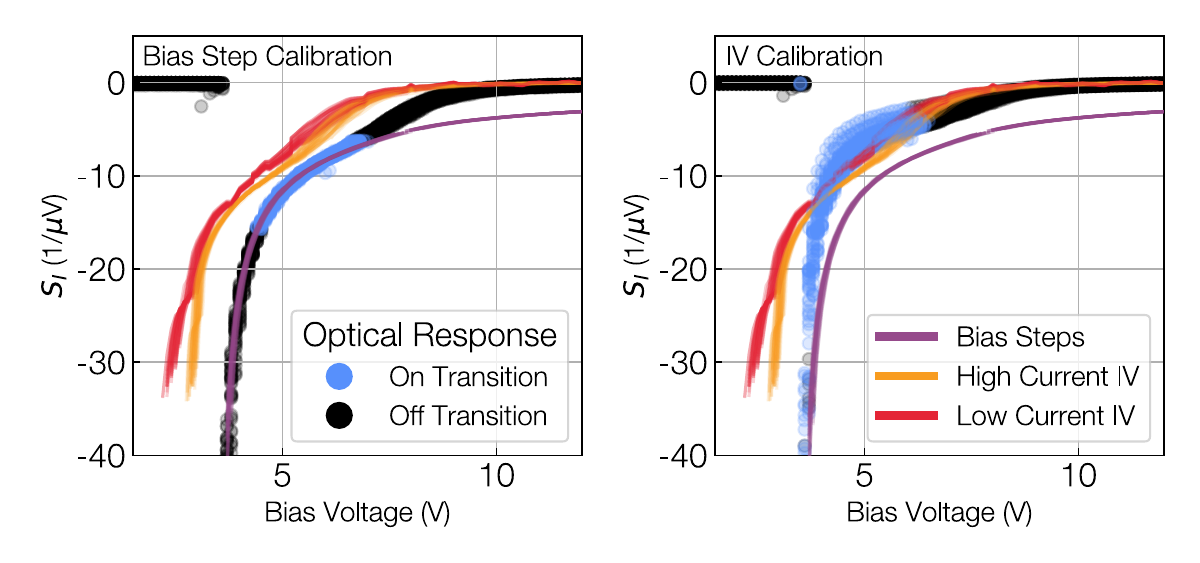}
    \caption{\label{fig:gain_cal} The comparison between different detector responsivity calibration methods and the observed optical response to a chopped source as a function of commanded bias voltage, where the bias line resistance is 16.4~$\mathrm{k\Omega}$. In both panels, the responsivity curves are shown for 39 detector channels at 90~GHz, all close to 50\% $R_n$ at $V_\mathrm{bias}=6$~V. The lines indicate the $S_I (V_\mathrm{bias})$ values calculated using either the IV curves or bias steps method. In the left panel, bias step calibrations are used to determine the range of bias voltages ``on transition'' between 20-80\% $R_n$, and this range is fit to the $S_I$ function. The same operation is performed in the right panel for the ``high current'' IV calibration. The bias step calibration method correctly reproduces the optical response pattern when the detectors are on-transition while the IV calibration does not follow the observed optical response. This is believed to be due to non-trivial wafer heating that occurs during an IV curve measurement.}
\end{figure*}

Converting the electrical signal from the readout system into the meaningful quantity of optical power requires a careful calibration process. Detector responsivity refers to the calibration factor between the change in incident optical power on a bolometer and the corresponding change in current through that bolometer. Calibrating this responsivity is necessary for on-sky observations and in-lab measurements where the analysis is not relative to a normalized value. For in-lab measurements in particular, the detector responsivity is critical to constraining the end-to-end efficiency of the optical system, such as in Section~\ref{subsec:opt_resp}. 

Raw data from the readout system is recorded in phase units that represent the phase shift of a resonator in response to changes in flux~\citep{Yu2022_SMURF}. This phase is proportional to the current through the bolometer, meaning that the phase can be converted to current based on the configuration of the readout system at the time that the data was taken. In practice, observations with our detectors produce time ordered data representing the measured change in current through the voltage-biased TES bolometer. The responsivity calibration is used to convert this quantity into a change in optical power on the detectors.

The responsivity of a TES bolometer is a function of many parameters, including the bath temperature, electrical bias power ($P_\mathrm{bias}$) and current ($I_\mathrm{bias}$), as well as the total optical power on the detector ($P_{\gamma}$). The typical methods for calibrating these detectors use either IV (current-voltage) curves or bias steps at discrete bias points to calculate the detector response~\citep{appel2022}. The comparison of these two methods is presented in this section using the Simons Observatory detector arrays. The performance of each method is evaluated within the context of calibrating detector timestreams and measuring the optical efficiency of the end-to-end optics tube.

Following the discussion in~\cite{appel2022}, for detectors on-transition, the total power through a TES bolometer is

\begin{equation}
    P_\mathrm{tot} \equiv P_{sat} = P_\mathrm{bias} + P_\mathrm{\gamma}.
\end{equation}

\noindent $P_{sat}$ is a function of the bath temperature and detector properties such as the thermal conductivity between the bolometer island and the thermal bath. In equilibrium, $P_{sat}$ is constant; changes in the incident optical power are canceled by changes in the electrical power through the detectors as

\begin{equation}
    dP_\mathrm{tot} = dP_\mathrm{bias}(I_\mathrm{bias}, I_\mathrm{TES}) + dP_\mathrm{\gamma}(I_\mathrm{bias}, I_\mathrm{TES}) \ .
\end{equation}

\noindent The optical response of the detectors, $S_\gamma$, is the change in current measured through the bolometer for a change in incident optical power, assuming constant bias power. Under a set of equilibrium conditions that include constant bath temperature, constant optical loading and strong negative electrothermal feedback, this quantity is equivalent to $S_I$, the detector current response to a changing bias power. These can be expressed mathematically as

\begin{align}
    S_\gamma &=  \frac{\partial I_\mathrm{TES}}{\partial P_\mathrm{\gamma}} \bigg|_{I_\mathrm{bias} = I_0} \\
    S_I &=  \frac{\partial I_\mathrm{TES}}{\partial P_\mathrm{bias}}\bigg|_{I_\mathrm{bias} = I_0} \ .
\end{align}

\noindent Therefore, IV curves or bias steps taken under $S_\gamma = S_I$ conditions can be used to calculate the optical responsivity, $S_I$, of a detector. The inverse, $S_{I}^{-1}$, is the measured calibration factor for converting the change in bolometer current into the change in optical power on the bolometer.

As discussed in~\cite{wang2022_ufm_bias}, wafer heating in the Simons Observatory focal-plane modules has been shown to impact the accuracy of using $S_I$ to predict $S_\gamma$ with IV curves and the level of this effect depends on the exact parameters used while taking the IV curves. The impact of the wafer heating effect for IVs taken following the methods planned for operations was tested with the LATR-Tester using a chopped optical source of a known constant temperature. Using this setup, we verify that the responsivity predicted by bias steps is not similarly impacted by wafer heating.

The procedure for comparing responsivity calibration methods is as follows. First, with the source off, IV curves were recorded using two different parameter sets under consideration as the default IV method for Simons Observatory observations. The parameter sets differ in the amount of bias current applied during an initial overbiasing step of the IV curve. One set uses a higher current (``high current IV'') while the other uses a lower current intended to reduce wafer heating (``low current IV''). In both cases, the initial overbiasing step still induced heating on the focal plane, so the temperature is given sufficient time to return to its pre-IV temperature between each operation. The results of these IVs are used to calculate the $S_I(I_{bias})$ values expected from this calibration method. 

Next, the bias voltage is slowly stepped down from 12~V to 0~V, with bias step calibrations and chopped source observations taken at each bias point. Care is taken to allow the wafer to thermalize after each bias point change. 
The observed change in current in response to the chopped source is
\begin{equation}
    \Delta I_\mathrm{det}(I_\mathrm{bias}) =  \eta_\mathrm{det} \Delta P_\mathrm{source} \mathcal{S_\gamma}(I_\mathrm{bias}) \ ,
\end{equation}
\noindent where $\Delta P_\mathrm{source}$ is the chopped in-band optical power of the source and $\eta_\mathrm{det}$ accounts for the individual detector's coupling to the source. 

Since $\eta_\mathrm{det}\Delta P_\mathrm{source}$ is constant for the duration, $\Delta I_\mathrm{det}(I_\mathrm{bias})$ is template fit to the $S_I(I_\mathrm{bias})$ curves measured for each responsivity calibration method under test over the range where the calibration method finds the detectors to be ``on-transition'' (between 20-80\% $R_n$, normal resistance). Figure~\ref{fig:gain_cal} shows the results of these fits for bias steps and the high current IV calibrations for a subset of detectors found to be close to 50\% $R_n$ at $V_\mathrm{bias} = 6$~V. The bias step calibration method predicts a responsivity that matches the observed optical response for the range where the detectors are found on-transition and expected to be optically responsive. The IV calibrations do not predict the observed pattern in the optical response and, assuming the bias step calibrations are correct, would misestimate the responsivity calibration factor by about a factor of 2 while detectors are on transition. These results validate the decision to baseline the use of bias steps to calibrate the optical responses of Simons Observatory detector arrays.

\section{\label{sec:efficiency} Optical Efficiency}
\subsection{\label{sec:cold_load} Detector Array Efficiency}
To obtain optical efficiencies of the detectors, the LATR-Tester is placed in its ``dark'' configuration with a metal plate replacing the window and a cold load installed at the end of the 4~K optics. The cold load is a plate covered with flat metamaterial absorptive tiles with the same design as those in~\cite{Xu_2021}. The optics tube design means the cold load face is within 5~mm of a secondary focus in the optics. When combined with the cold load dimensions, it is expected that the load is completely beam filling for all pixels on the detector array. The cold load plate is made of 1100 aluminum to maximize thermal conductivity. A 250~$\Omega$ heater was mounted on the back of the plate. A thermometer diode is mounted beside the heater and another at the edge of the cold load to monitor the temperature gradient during the measurements.

Since the cold load completely fills the beam of the detectors, it can be assumed that the cold load temperature is also the temperature of the signal seen by the detectors. By changing the cold load temperature $T_\mathrm{CL}$, the change in incoming optical power $\Delta P_\mathrm{opt}$ can be recorded for each detector in the form of the electrical power necessary to bias that detector:
\begin{equation}
    \Delta P_\mathrm{opt} = P_\mathrm{bias}(T_\mathrm{CL}) - P_\mathrm{bias}(T_0) \ ,
\end{equation}

\noindent where $P_\mathrm{bias}(T_0)$ is the bias power at the lowest cold load temperature, typically 7~K. The electrical bias power at each step is recorded using IV curves. $P_\mathrm{bias}$ at each cold load temperature is taken to be the applied bias power that drives the detectors to 90\% normal resistance. Knowing both the input temperature and received power allows for efficiency of the optical system to be estimated as
\begin{equation}
    \eta = \frac{\Delta P_\mathrm{opt}}{\Delta P_\mathrm{exp}} \label{eq:eta} \ ,
\end{equation}

\noindent where $\Delta P_\mathrm{exp}$ is the expected optical power change on the detectors over the course of a cold load temperature sweep. At any fixed $T_\mathrm{CL}$, the expected detector loading is

\begin{equation}
    \label{eq:p_exp}
    P_\textrm{exp} = \int^\infty_0 f(\nu) \Big( p_\mathrm{CL}(\nu, T_\mathrm{CL}) + p_\mathrm{LS}(\nu,T_\mathrm{LS}) \Big)  d\nu \,
\end{equation}
\noindent where $p_\mathrm{CL}$ and $p_\mathrm{LS}$ are the Planck emission profiles of the cold load and Lyot stop at their respective temperatures, and $f(\nu)$ is the transmission response of the instrument. The observed change in Lyot stop temperature during this sweep, $dT_\mathrm{LS}/dT_\mathrm{CL}$, produces a small but non-zero effect on the change in expected detector loading.

To correct for thermal effects, we perform a dark subtraction using dedicated dark pixels on the array that have no orthomode transducer to allow optical coupling through the on-chip circuitry. The bolometers for these dark pixels are designed with a lower thermal conductance $G$ than the optical bolometers, so a scaling factor $\chi = G_\mathrm{dark}/G_\mathrm{opt}=0.55$ is required to perform a dark subtraction between the two bolometer types. This scaling factor assumes the same critical temperature between dark and optical bolometers, an assumption which has been shown to be valid with these detector arrays~\citep{Dutcher_2023}. The dark-corrected optical power is expressed as
\begin{equation}
    \Delta P_\mathrm{opt} = \Delta P^\mathrm{raw}_\mathrm{opt} - \frac{1}{\chi} \Delta P_\mathrm{dark} \ ,
\end{equation}
\noindent where $\Delta P_\mathrm{dark}$ is the average response of the dark detectors to the changing cold load, and $\Delta P^\mathrm{raw}_\mathrm{opt}$ is the response of the optically-coupled detectors without any dark correction. This dark correction leads to a $\sim10\%$ reduction in the measured optical efficiencies of the detectors.

If no assumptions are made about the instrument passband $f(\nu)$, except that it is a tophat response with 100\% transmission, then Equation~\ref{eq:eta} yields an optical efficiency measurement for the end-to-end optics tube. This includes the efficiency of the detectors and all optical elements at 4~K and below, but not including any of the frontend filtering at the warmer stages of the instrument. This internal optics tube efficiency comes out to be 21\% / 26\% (90 / 150 GHz). These numbers exceed the optical efficiency expected for these components from the baseline instrument model and is therefore a powerful indicator that there are no significant issues with the detectors or optics at 4~K and below.

To extract a detector-only efficiency, we must also account for loss through the optical elements between the cold load at 4~K and the detector array at 100~mK. This includes the Lyot stop, a polypropylene IR blocking filter, 3 silicon lenses, and 2-3 low pass filters (see Section~\ref{sect:holog-method}). The transmission profiles of these optics are characterized through a combination of in-lab measurements and simulations. In total, we expect a transmission loss of about 79\% / 56\% (90 / 150~GHz) through the internal optics out to 4~K. This loss is largely driven by the Lyot stop, which was designed to control excess loading at the cost of significant beam truncation. For this efficiency measurement, we assume a transmission loss at the Lyot stop of about 76\% / 49\% (90 / 150~GHz), as expected from beam simulations. The transmission responses of these optical components are combined into the expected instrument passband, $f(\nu)$ in Equation \ref{eq:p_exp}, to divide out the effects of the sub-4~K optics and obtain a detector-only optical efficiency. We define detector efficiency as a measurement of the transmission through the feedhorn, orthomode transducer, and on-chip electronics.

\begin{figure}
    \centering
    \includegraphics[width=\linewidth]{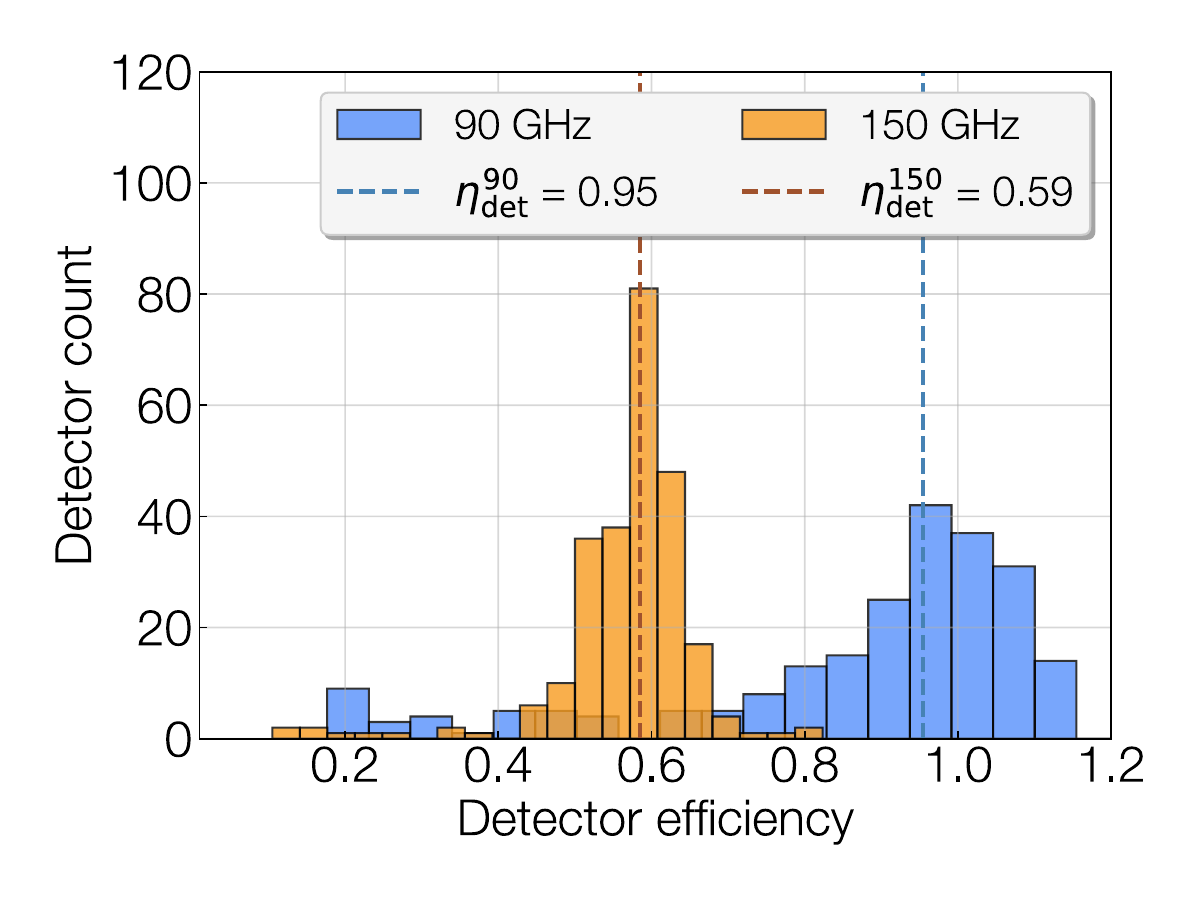}
    \caption{Detector optical efficiencies measured using a cold load in the LATR-Tester's dark configuration. These values assume a transmission model for all optics between 4~K and 100~mK. The median efficiency values of the distribution are 95\% and 59\% for 90 and 150~GHz detectors, respectively. Detectors with measured efficiencies above unity point to a systematic error in this measurement, likely from uncertainty in the optics model or excessive wafer heating.}
    \label{fig:dark_etas}
\end{figure}

Figure~\ref{fig:dark_etas} shows the detector efficiencies for all operating detectors on this array with median values of 95\% / 59\% (90 / 150~GHz). These numbers are consistent with previous dark measurements performed on the same detector array, with a median-value agreement to within 15\%. We note that the efficiencies for the 90 GHz detectors appear high, with many channels measuring above 100\%. Several factors could contribute to this. An inaccurate transmission model of the 4~K optics that are corrected for when calculating the expected optical loading is one likely culprit. If the loss assumed for the low pass filters or silicon lenses is too low, for example, that would lead to an optical efficiency that is biased high. Excess wafer heating would also have an effect on the observed biasing power, although attempts were made to correct for this.

\subsection{\label{sec:warm_load} Integrated Optics Tube Efficiency}

\begin{figure*}[t]
    \centering
    \includegraphics[width=\textwidth]{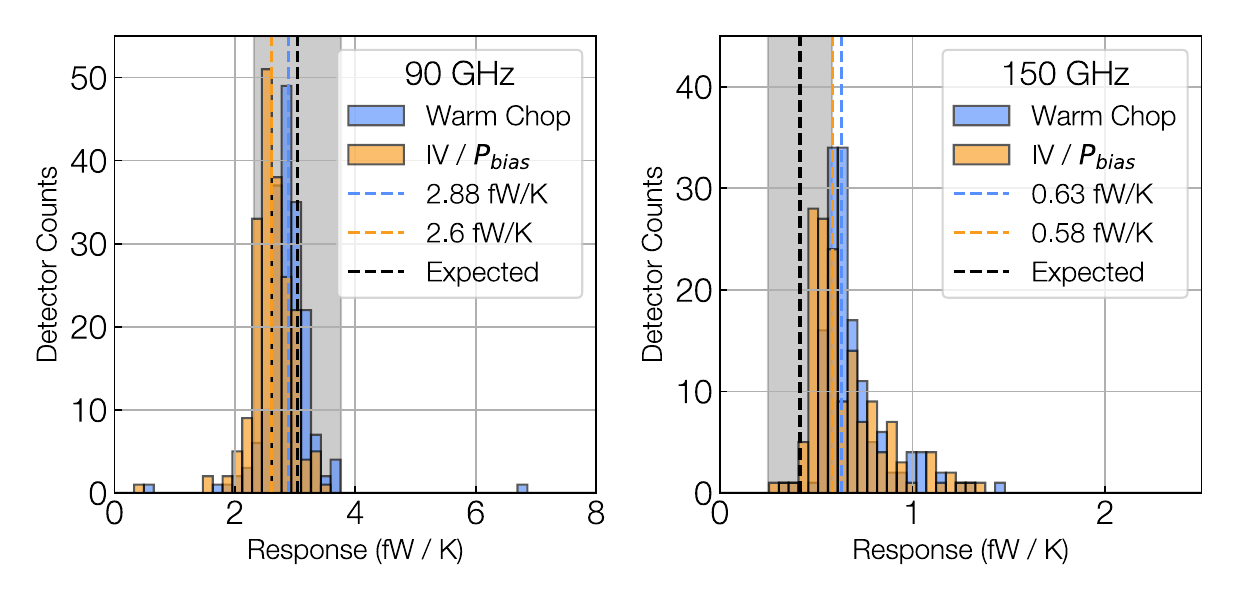}
    \caption{\label{fig:optical_resp}  The measured optical responses for the 90~and 150~GHz spectral bands found using the two different measurement methods, ``warm chops'' and ``IV / $P_\mathrm{bias}$,'' as explained in the text. The measured optical responses of the 150~GHz detectors are smaller than that of the 90~GHz detectors because of the lower NDF transmission in the 150~GHz band. The colored dashed lines mark the median values measured for each method. The black dashed lines mark the response expected from the end-to-end optics model for the installed optics and detectors. The gray shaded region indicates the effect of the band-averaged NDF transmission uncertainty on the expected optical response. Both measurement methods are in good agreement with the optical model, indicating the integrated optical efficiency is sufficiently close to the designed values.}
\end{figure*}

The integrated optical efficiency, combining efficiencies of the detector array, Lyot stop, reimaging optics, and frontend filtering, is proportional to the final mapping speed of the instrument. Thus, it is important to validate that the integrated design matches the expected values. 

\label{subsec:opt_resp}

Optical response measurements were performed using two different methods, ``warm chop'' and ``IV / $P_\mathrm{bias}$.'' In the warm chop method, a temperature-controlled beam-filling blackbody source, $T_1$, is mounted above the cryostat window. The temperature of $T_1$ is stepped through a range of temperatures between $10^\circ$C and $40^\circ$C. At each point, IV and bias steps calibrations are performed as described in Section~\ref{sec:detector-gain} during the process of biasing the detectors onto transition, with sufficient thermal settling time between each operation. Then a room-temperature beam-filling source, $T_2$, is chopped in front of $T_1$ several times while data is streamed from the detectors. The temperature of $T_1$ is read out concurrently with measurement using nine 10~$\mathrm{k\Omega}$~NTC thermistors mounted on the back of the source and the temperature of $T_2$ was measured using an IR thermometer before and after each measurement. 

Each $T_1$ temperature setting is used to measure the change in power observed for a specific $\Delta T = T_1-T_2$, where the detector timestreams are calibrated into power using the bias step responsivity calibration. The measurements for each detector are fit to a line where the slope is then the per-detector optical response. A non-zero y-intercept in these measurements indicated some unaccounted for difference in spill between the two sources or a differential emissivity between the two. Both effects are calibrated out by forcing the y-intercept to zero and appropriately adjusting the slopes of the lines to match. This self-calibration causes a less than 0.5\% shift in the median optical response of the array. In addition, the measured response of the non-optically coupled, ``dark,'' detectors is used to remove any non-optical pickup during the measurement; this is a $\sim1\%$ effect for the 90~GHz detectors and a $\sim5\%$ effect for the 150~GHz detectors. The distribution of optical responses for the array are shown in Figure~\ref{fig:optical_resp}, which includes careful detector cuts to ensure non-responsive detectors or detectors with spill around the NDF do not contaminate the sample. The statistical errors for the response of each detector are well below the spread seen in the array. 

The optical response was also measured using the ``IV / $P_\mathrm{bias}$'' method. This method is analogous the one used in the cold load detector efficiency measurements, where the bias power necessary to bias the detectors to 90\% normal resistance is calculated as a function of optical load using IV curve measurements. IVs were taken at the same range of temperatures of the source $T_1$ during the warm load chop; these are combined with IVs taken when a 77~K liquid nitrogen source is placed in front of the window to achieve a $\Delta T \sim 215$~K difference to observe a change in $P_\mathrm{bias}$ as a function of input temperature. Linear fits to the change in $P_\mathrm{bias}$ power as a function of input temperature are used to measure the per-detector responses with this method. 

As with the warm chop method, dark detectors are used to constrain and remove any non-optical response during these measurements. This method had significantly higher non-optical response when compared to the warm chop method, with the dark subtraction correction reducing the response by $18\%$ and $70\%$ for the 90 and 150~GHz detectors, respectively. The dark subtracted optical responses for the IV / $P_\mathrm{bias}$ measurements are also shown in Figure~\ref{fig:optical_resp} and are in good agreement with the warm chop method.

The black dashed lines in Figure~\ref{fig:optical_resp} indicate the expected optical response for each band using an integrated optical model that includes the transmission spectra of the various optical elements, the bandpasses and detector efficiencies measured in previous sections, and the calibrated NDF transmission. The gray shaded region marks the uncertainty on the NDF transmission, while this is not the only source of uncertainty in the model it is believed to be the largest one. 

The integrated optical model expects the end-to-end efficiencies of the 90~and 150~GHz bands to be 16.4\% and 26.4\%, respectively. Combining the two measurements and accounting for the NDF uncertainty predicts measured efficiencies of $14.9\pm3.7\%$ and $38.9\pm16.9\%$. Both measurements are within one-sigma of their expected values with the error dominated by the NDF transmission uncertainties. This agreement indicates there are no substantial issues with the transmission of the different optical elements and that the optical model, where the largest reduction in efficiency comes from the designed Lyot stop efficiency, can accurately predict the integrated optical efficiency of the system.

\section{Predicted On-Sky Performance} \label{sec:on-sky}
The expected instrument performance of the Simons Observatory has been closely tracked since the project began in 2017. Sensitivity models of both the large- and small-aperture telescopes are compiled using BoloCalc~\citep{hill2018bolocalc}, a highly-detailed sensitivity calculator for CMB experiments. These models incorporate an array of inputs that include the transmission and thermal loading of every component in the optics chain, detector properties, exact bandpass shapes, and the expected atmosphere impact from the Chilean site. As new data becomes available, e.g., through updated designs or in-lab characterizations, the instrument model and its predicted sensitivity becomes more robust.

Contributions to detector noise estimates in this calculation include the thermal carrier noise in the bolometers, readout noise from the superconducting amplifiers, and photon noise from the both the sky and the instrument itself. Converting this to a noise-equivalent CMB temperature (NET) yields a useful performance metric that quantifies the ability of the instrument to measure fluctuations in the CMB temperature.

The LATR-Tester characterization of the mid-frequency optics tube outlined in this work comprise the final and most rigorous update to the large aperture telescope forecast model. These model updates include the measured end-to-end passband (Section \ref{sec:bandpass}), beam spillover to 300~K (Section \ref{sec:beams}), and optical efficiency of the detector arrays (Section \ref{sec:efficiency}). From these measurements, the on-sky NET of the large aperture telescope's 90 and 150~GHz channels are predicted to be $3.3\pm0.7$ and $3.9\pm0.9$ $\mathrm{\mu K \sqrt{s}}$, respectively, in the nominal configuration with four mid-frequency optics tubes. This sensitivity forecast exceeds the baseline performance requirements~\citep{Ade_2019} with an NET margin of over 30\% in the worst case scenario for each spectral band, giving strong indication that the instrument will achieve its goals on deployment.

\section{Conclusions}

The LATR-Tester program has demonstrated that the Simons Observatory large aperture telescope 90/150~GHz camera is expected to perform at a level that exceeds the requirements set by the Simons Observatory science goals at the outset of the project. This is an integral step in the deployment of the instrument to ensure the most efficient observation time. The testing plan was built on the heritage of past CMB projects to leverage measurement techniques and to highlight critical areas of importance. In particular, we draw on lessons learned from several generations of cameras on the Atacama Cosmology Telescope (ACT), upon which much of the optics tube design is based on. The holography measurement, for example, was developed in response to optical anomalies observed in ACT's far-field beams. The development of the holography method has validated the far-field response of this optics tube and led to a tight constraint on ambient light spillover, which is one of the primary drivers of sensitivity in the instrument. Additionally, it has led to the removal of a filter from the nominal optics tube design that would have degraded beam quality and mapping speed in the 90~GHz channel. Thermal beam maps support these results as being representative of the expected performance on the deployed instrument. 

Spectral measurements across the detector array at every frequency region of interest from 1 to 1000~GHz confirm that the end-to-end passband targets are met and show that there are no significant out-of-band leaks that would degrade mapping performance. The polarization performance of the system was tested with separate independent measurements that place strong limits on the cross-polarization impact from both the optics and the detectors. We validate two important detector metrics which can only be tested in a fully-integrated system like the LATR-Tester: long-term readout stability in an optical environment and optical time constants. Finally, we characterize the optical efficiency of the detector array with an internal cold load, and the optical efficiency of the integrated optics tube with an external warm load.

Altogether, we have shown viable procedures for the in-lab testing of virtually every performance metric that significantly impacts the sensitivity and understanding of our optics tubes. These tests provide an optimistic picture of the performance expected on the large aperture telescope's 90 and 150~GHz channels when it is deployed and observing the sky in 2024. Prior to deployment, similar testing plans will be performed for the low-frequency 30 / 40~GHz and ultra-high-frequency 220 / 280~GHz optics tubes to validate their performance as well. Given the scope and scale of modern-day CMB experiments, integrated testing plans like the one outlined in this paper are expected to become the new standard.
\\


\begin{acknowledgments}
This work was supported in part by a grant from the Simons Foundation (Award \#457687, B.K.).
\end{acknowledgments}

\bibliography{LATRtPaper}
\bibliographystyle{aasjournal}

\end{document}